\PassOptionsToPackage{hidelinks}{hyperref}
\documentclass[aps,superscriptaddress,twocolumn,pra,raggedbottom]{revtex4-2}
\usepackage{orcidlink}
\usepackage{soul}
\usepackage{}
\usepackage[normalem]{ulem} % GB review: \sout is available; normal emphasis is preserved.
\usepackage{dsfont}
\usepackage{tikz}
\usetikzlibrary{decorations.pathmorphing}
\usepackage[utf8]{inputenc}
\usepackage{physics}
\usepackage{braket}
\usepackage{amsmath}
\usepackage{amssymb}
\usepackage{placeins}
\usepackage{dsfont}
\usepackage{appendix}
\usepackage{xcolor}
\usepackage{graphicx,bbm}
\usepackage{comment}
\usepackage[hidelinks]{hyperref}
\hypersetup{
	colorlinks,
	linkcolor={red!50!black},
	citecolor={blue!50!black},
	urlcolor={blue!80!black}
}
\usepackage{mathtools}

\begin{document}
% GB REVIEW COPY: original scientific text and displayed equations retained.
% Blue annotations are proposals/questions, not adopted scientific changes.
% Short blue strikeouts mark proposed deletions; no section has been removed.

\title{Non-Markovian exceptional points and information backflow: \\a Heisenberg reaction-coordinate approach}
\author{Daniel Holst\,\orcidlink{0009-0007-4875-2454}}
  \email{daniel.holst@fysik.lu.se}
  \affiliation{Physics Department and NanoLund, Lund University, Box 118, 22100 Lund, Sweden}
\author{Gianmichele Blasi\,\orcidlink{0000-0002-4024-788X}}
  \affiliation{Instituto de F\'isica Interdisciplinar y Sistemas Complejos IFISC (CSIC-UIB), E-07122 Palma de Mallorca, Spain}
\author{Shishir Khandelwal\,\orcidlink{0000-0003-3366-624X}}
\email{shishir.khandelwal@fysik.lu.se}
  \affiliation{Physics Department and NanoLund, Lund University, Box 118, 22100 Lund, Sweden}

\begin{abstract}
   The primary pathway to the investigation of exceptional points (EPs) in open quantum systems has been the use of Markovian master equations. However, recently it has been realised that they can arise in settings that show non-Markovian behaviour. The main approach in this direction has been the use of pseudomodes, which allow the influence of a structured environment to be replaced by new damped auxiliary modes. In this work, we use a Heisenberg-equation approach which relies on the reaction-coordinate formalism to obtain exact analytical dynamics. First, we show how our method considerably simplifies the analysis compared to the use of master equations. Second, we utilise it to discuss non-Markovian revivals in systems with second and higher-order EPs, showing that the notion that EPs demarcate the onset of revivals is not generally true. We then utilise our results to discuss the dynamical effects of higher-order non-Markovian EPs, deriving bounds on the number of revivals and investigating enhanced asymptotic relaxation near higher-order EPs. Finally, we investigate non-Markovian EPs with engineered reservoirs and show how they can be utilised to control the rate of irreversible information loss. 

\end{abstract}

\maketitle

\section{Introduction}
Exceptional points (EPs) are the paradigmatic property of non-Hermitian systems. While they were first theorised and experimentally realised in classical and semi-classical systems \cite{Heiss_1990,Dembowski2001,Heiss_2012,El-Ganainy2018,Miri2019}, EPs have now long been studied in quantum settings and have been shown to give rise to various distinctive and potentially useful phenomena, such as improving the performance of quantum heat engines, aiding asymmetric mode conversion \cite{Chen2022,Kumar2021,Sun2024,Gao2025,svegborn2026}, optimizing relaxation \cite{Khandelwal2021,Zhang2021a,zhang2022,Zhou2023,Kumar2022a}, generating entanglement \cite{Khandelwal2024a,Li2023,Teixeira2023,Yu2025}, and assisting quantum-transport phenomena \cite{Saha2023,Bourgeois2026Tomography}. The conventional approach in this direction has been the use of Markovian master equations as they can naturally be expressed in the form of non-Hermitian matrices. The resulting EPs have commonly been called \textit{Liouvillian EPs} \cite{Minganti2019}.

While Liouvillian EPs are generally understood to be a property of Markovian master equations, recent works have shown two key results. First, it has been demonstrated that ordinary Liouvillian EPs can also be seen in the exact dynamics of a system. In particular, it has been established that they rely on the Markovianity of the dynamics, and not necessarily on an assumption about the strength of the system--bath coupling \cite{Khandelwal2025em}. Second, it has been shown that a distinct class of EPs can arise in non-Markovian approaches, specifically ones involving pseudomodes, in which the dynamics of a system is inferred from the dynamics of a larger system including effective degrees of freedom \cite{deVega2017}. In the simplest models, these EPs have been shown to disappear in the Markovian limit and mark a boundary between Markovian and non-Markovian dynamics. Therefore, they have been referred to as \textit{Non-Markovian EPs} \cite{Lin2025,Zhang2025,Wong2026,Zhang20261,Longhi2026,Sergeev,zhang2025expl,Shen_2026,Ghosh2026}. The key idea behind non-Markovian EPs stems from the fact that the Markovian property of the evolution of a large system is not preserved when looking at its subsystems. For example, a system may be evolving according to a Markovian master equation; however, the reduced dynamics of a part of this system are not necessarily Markovian. In this picture, the remaining degrees of freedom store memory of the subsystem’s past evolution, and the non-Markovian EP governs their coupled dynamics.
\\
The goals of this work are manifold. We first introduce a pathway to investigate non-Markovian EPs in the Heisenberg equations of motion through the reaction-coordinate formalism \cite{Strasberg2016,nazir2018}, demonstrating how this approach considerably simplifies the analysis compared to master equations by restricting the analysis to a physically relevant subspace of the dynamics. We then shift our focus to analytically solvable examples of second- and higher-order EPs, showing the exact conditions needed to reach them. Next, we investigate the connection between EPs and non-Markovian revivals. Previous works \cite{Lin2025,Longhi2026} have shown that second-order EPs mark the onset of non-Markovian revivals. We show that the coincidence between EPs and the onset of information backflow is
configuration dependent; specifically for higher orders, EP curves need not coincide with the backflow boundary. Next, we investigate the dynamical consequences of EPs, deriving bounds on the number of revivals and showing how dynamics are sped up at higher-order EPs. Finally, since most of the literature on non-Markovian EPs has relied on the assumption of Lorentzian spectral densities, as our final result, we investigate EPs in engineered, non-Lorentzian scenarios. For such an example, we characterize when a particle escapes into the environment and show how tuning to an EP determines the escape-time distribution and optimizes the long-time escape rate.

\section{General scenario: reaction-coordinate formalism}
\label{sec:gen}

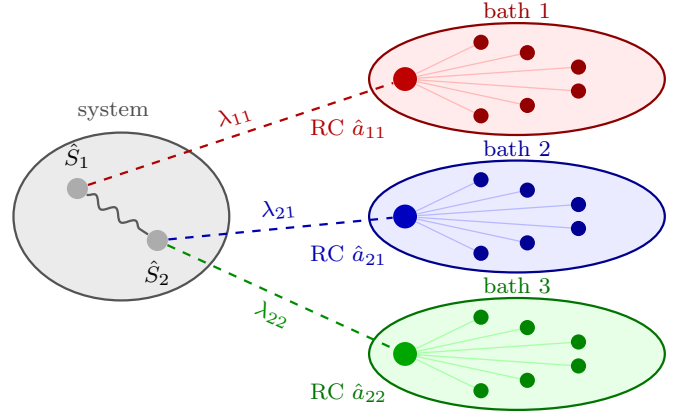
\begin{figure}[t]
\centering
\resizebox{\columnwidth}{!}{%
\begin{tikzpicture}[
    font=\footnotesize,
    sysmode/.style={circle,draw=none,fill=gray!65,minimum size=7.5pt,inner sep=0pt},
    bathmode/.style={circle,draw=none,minimum size=5.4pt,inner sep=0pt},
    rcmode/.style={circle,draw=none,minimum size=8.5pt,inner sep=0pt},
    rcconnect/.style={thick,dashed},
    residualconnect/.style={line width=0.45pt},
    intconnect/.style={thick,gray!70!black,decorate,decoration={snake,amplitude=0.6mm,segment length=3mm}}
]
    % System and modes
    \draw[fill=gray!16,draw=gray!65!black,thick] (0.1,0) ellipse[x radius=1.35,y radius=1.05];
    \node[gray!60!black] at (0,1.32) {system};
    \node[sysmode,label=above:{$\hat S_1$}] (s1) at (-0.45,0.35) {};
    \node[sysmode,label=below:{$\hat S_2$}] (s2) at (0.55,-0.3) {};
    \draw[intconnect] (s1) -- (s2);

    % bath 1
    \draw[fill=red!8,draw=red!55!black,thick] (5.05,1.72) ellipse[x radius=1.85,y radius=0.70];
    \node[red!60!black] at (5.05,2.58) {bath 1};
    \node[rcmode,fill=red!75!black,label distance=1pt,label={[red!70!black,yshift=-8pt]below left:{RC $\hat a_{11}$}}] (rcrc) at (3.65,1.72) {};
    \foreach \x/\y in {4.60/2.18,5.18/2.05,5.82/1.87,5.82/1.57,5.18/1.39,4.60/1.26} {
        \draw[residualconnect,red!28] (rcrc) -- (\x,\y);
        \node[bathmode,fill=red!58!black] at (\x,\y) {};
    }

    % bath 2
    \draw[fill=blue!8,draw=blue!55!black,thick] (5.05,0) ellipse[x radius=1.85,y radius=0.70];
    \node[blue!60!black] at (5.05,0.86) {bath 2};
    \node[rcmode,fill=blue!75!black,label distance=1pt,label={[blue!70!black,yshift=-4pt]below left:{RC $\hat a_{21}$}}] (bcrc) at (3.65,0) {};
    \foreach \x/\y in {4.60/0.46,5.18/0.33,5.82/0.15,5.82/-0.15,5.18/-0.33,4.60/-0.46} {
        \draw[residualconnect,blue!28] (bcrc) -- (\x,\y);
        \node[bathmode,fill=blue!58!black] at (\x,\y) {};
    }

    % bath 3
    \draw[fill=green!10,draw=green!45!black,thick] (5.05,-1.72) ellipse[x radius=1.85,y radius=0.70];
    \node[green!45!black] at (5.05,-0.86) {bath 3};
    \node[rcmode,fill=green!65!black,label distance=1pt,label={[green!45!black,yshift=-4pt]below left:{RC $\hat a_{22}$}}] (gcrc) at (3.65,-1.72) {};
    \foreach \x/\y in {4.60/-1.26,5.18/-1.39,5.82/-1.57,5.82/-1.87,5.18/-2.05,4.60/-2.18} {
        \draw[residualconnect,green!35] (gcrc) -- (\x,\y);
        \node[bathmode,fill=green!52!black] at (\x,\y) {};
    }

    % System-bath couplings to the reaction coordinates
    \draw[rcconnect,red!70!black] (s1) -- node[above,sloped] {$\lambda_{11}$} (rcrc);
    \draw[rcconnect,blue!70!black] (s2) -- node[above,sloped] {$\lambda_{21}$} (bcrc);
    \draw[rcconnect,green!55!black] (s2) -- node[below,sloped] {$\lambda_{22}$} (gcrc);

\end{tikzpicture}}
\caption{Example of the system--bath geometry used for the reaction-coordinate mapping, with $N_s=2$, $N_b^1=1$, and $N_b^2=2$.}
\label{fig:system-bath-geometry}
\end{figure}

We consider a general system with $N_s$ modes $\{\hat S_k\}_{k=1}^{N_s}$, with the $k$-th mode coupled with $N_b^k$ non-interacting baths. The total number of baths coupled with the system is therefore $N_b = \sum_{k=1}^{N_s}N_b^k$. A representative instance of this geometry, with two system modes and three baths, is shown in Fig. \ref{fig:system-bath-geometry}. The total Hamiltonian is given by $\hat H = \hat H_{\text{S}}+\hat H_{\text{B}} + \hat H_{\text{SB}}$, where 
\begin{equation}\label{eq:mod}
\begin{aligned}
    &\hat H_\text{B} = \sum_{k=1}^{N_\text{s}}\sum_{j=1}^{N_b^k}\sum_{\alpha} \omega_{kj\alpha}\hat b^{\dagger}_{kj\alpha}\hat b_{kj\alpha}, \\
    &\hspace*{0.3cm}\hat H_\text{SB} = \sum_{k=1}^{N_\text{s}}\sum_{j=1}^{N_b^k}(\hat S^\dagger_{k} \hat B_{kj} + B_{kj}^\dagger\hat S_{k}),
\end{aligned}
\end{equation}with system operators $\hat S_k$ and collective bath operators $\hat B_{kj} = \sum_\alpha t_{kj\alpha} \hat b_{kj\alpha}$. Here, $k$ indexes the system mode, $j$ indexes the corresponding baths, and $\alpha$ indexes individual modes of the baths. The baths can be fermionic or bosonic and the system Hamiltonian is in principle arbitrary and can contain interactions and non-linearities. For simplicity, we have assumed that distinct system operators are coupled to distinct baths, as this is relevant for our examples. Thus, a pair $(k,j)$ labels one independent channel: two different system modes do not share the same collective bath operator. For initially independent baths with zero first moments, the freely evolving bath operators satisfy $\expval{B^\dagger_{kj}(t)B_{k'j'}(0)}=0$ for $(k,j)\neq(k',j')$. Each bath has an associated damping kernel $K_{kj}$, defined by
\begin{align}
    K_{kj}(t) = \int \frac{d\omega}{2\pi} e^{-i\omega t} J_{kj}(\omega), 
\end{align}with the spectral density $J_{kj}(\omega) = 2\pi\sum_{\alpha}|t_{kj\alpha}|^2\delta(\omega-\omega_{kj\alpha})$. The overall memory kernels and the spectral densities for the modes are therefore $K_k(t) = \sum_{j} K_{kj}(t)$ and $J_k(\omega)=\sum_{j} J_{kj}(\omega)$.
In particular, consider a flat spectral density (i.e., constant $J_k(\omega) = \Gamma_k \hspace{2pt}  \forall k$ across the energy spectrum), for which the memory kernel reduces to a Dirac delta, $K_k(t) = \Gamma_k\delta(t)$ and has no temporal width. Such models can often be analytically and non-perturbatively solved \cite{Schaller2014,Blasi2024,Khandelwal2025em} for the evolution of the system. On the other hand, a structured spectral density can lead to long-lived bath correlations, giving rise to non-Markovian dynamics of the system \cite{Breuer2016}. While such scenarios have no general solution, the reaction-coordinate formalism \cite{Strasberg2016,nazir2018} can provide a considerable simplification. This approach incorporates collective bath degrees of freedom into the system, leaving a residual bath that may admit a simpler treatment. With a suitable choice, the residual bath becomes less structured, at the cost of an increased number of tracked degrees of freedom.

For spectral densities defined over all real frequencies, an analytically tractable class of structured baths is obtained when the memory kernel is of the form $K_k(t)= \sum_{j}\alpha^2_{kj} e^{-\delta_{kj} |t|-i\omega_{kj} t}$. Here $\omega_{kj}$ sets the bath resonance and $\delta_{kj}^{-1}$ is the decay timescale of the damping kernel. For $t\geq0$, this kernel is a sum of damped oscillations and corresponds to having Lorentzian spectral densities  
\begin{align}
      J_{kj}(\omega) = \kappa_{kj} \frac{\delta_{kj}^2}{(\omega-\omega_{kj})^2 + \delta_{kj}^2},
\end{align}where $\kappa_{kj}$ characterises the coupling strength between the system and bath $j$ of system mode $k$ and $\delta_{kj}$ sets the width of the spectral density, such that $\alpha_{kj} = \sqrt{\kappa_{kj}\delta_{kj}/2}$. This model can be systematically recast through reaction-coordinate mapping, which embeds the system into a larger one whose dynamics are local in time. This larger system includes the original system, as well as some modes of the bath, referred to as the reaction coordinates (RCs). As illustrated in Fig. \ref{fig:system-bath-geometry}, the mapping selects one collective mode from each bath as a reaction coordinate, while the remaining modes form residual baths. For the single-Lorentzian spectral density assumed for each independent
bath channel $(k,j)$, this construction introduces one RC per channel,
giving $N_b=\sum_k N_b^k$ RCs in total. Each RC couples only to its
corresponding system mode. It can be shown  \cite{nazir2018} that the effective model for the system--reaction coordinate-bath evolution is given by $\hat H_{\text{eff}} = \hat H_{\text S} + \hat H_{\text{RC}} + \hat H_{\text{S-RC}}+\hat{\tilde H}_{B} + \hat H_{\text{B-RC}}$, with the RC Hamiltonian $\hat H_{\text{RC}} = \sum_{kj}\Omega_{kj} \hat a_{kj}^\dagger \hat a_{kj}$, residual bath Hamiltonian $\hat{\tilde H}_{\text{B}} = \sum_{kj\beta }\Omega_{kj\beta}\hat{\tilde b}_{kj\beta}^\dagger\hat{\tilde b}_{kj\beta}$ and the interactions 
\begin{equation}
    \begin{aligned}
    &\hat H_{\text{S-RC}} =\sum_{kj} \lambda_{kj}(\hat S^\dagger_{k}\hat a_{kj} + \hat a^\dagger_{kj}\hat S_{k} )\\
    &\hat H_{\text{B-RC}} = \sum_{kj\beta} \left(\tilde t_{kj\beta} \hat a_{kj}^\dagger\hat{\tilde b}_{kj\beta} +\tilde t_{kj\beta}^* \hat{\tilde b}_{kj\beta}^\dagger\hat a_{kj}\right),
\end{aligned}
\end{equation}where $\Omega_{kj}=\omega_{kj}$ and $\lambda_{kj}=\sqrt{\kappa_{kj} \delta_{kj}/2}$. To distinguish the residual bath from the original one, we have used a $\sim$ and have changed the summation index $\alpha\to\beta$.  The transformed bath operators $\hat{\tilde{b}}_{kj\beta}$, as well as the RC operators $\hat a_{kj}$ satisfy the same (anti-)commutation relations as the original bath operators. Importantly, the effective spectral density corresponding to the bath-RC interaction is now flat, i.e., $\tilde J_{kj}(\omega) = 2\pi \sum_\beta |\tilde t_{kj\beta}|^2\delta(\omega - \Omega_{kj\beta})= 2\delta_{kj} $.  Thus, the RCs encode the finite memory of the original damping kernels. The joint system--RC operator equations have local
damping, while eliminating the RCs introduces a
finite-memory kernel for the original system.

\subsection{Equations of motion, exact dynamics, and exceptional points}
 We consider the Heisenberg equations for the evolution of system, RC and bath operators,  $\dot{\hat{o}}(t) = i[\hat H_{\text{eff}},\hat o(t)]$, with $\hat o = \hat S_k,\hat a_{kj}$ or $\hat{\tilde b}_{kj\beta}$. As shown in App. \ref{app:rc} (see also Refs. \cite{Blasi2024,Khandelwal2025em}), if the system Hamiltonian is quadratic and number conserving, the Heisenberg equations of the system and RC can equivalently be expressed as the following Langevin equation for the evolution of the $(N_s+N_b)$-dimensional vector 
$\vec v = (\hat S_{1},\dots,\hat S_{N_s},\hat a_{11},\dots,\hat a_{1 N^1_b},\hat a_{21},\dots)^T$, 
\begin{align}\label{eq:eomm}
    \dot{\vec v} = A\vec v + \vec \xi,
\end{align}where $\vec\xi = (0,\dots,0,\hat \xi_{11},\dots,\hat \xi_{1N_b^1},\hat \xi_{21},\dots,)^T$ with $\hat \xi_{kj} = -i \sum_\beta \tilde{t}_{kj\beta} e^{-i\Omega_{kj\beta} t}\hat{\tilde b}_{kj\beta}(0)$. $A$ is a $(N_s+N_b)\times (N_s+N_b)$ non-Hermitian matrix. The exact dynamics of the system and reaction coordinates can be determined by integrating the above,
\begin{align}\label{eq:eomsol}
    \vec v(t) = D(t) \vec v(0) + \int_0^t ds D(t-s)\vec \xi(s),\
\end{align}where $D(t)\coloneqq e^{At}$.  Using the above, the exact time-dependent reduced density matrix of the system--RC can in principle be calculated through expectation values of relevant system--RC operators.

 $A$ is a non-Hermitian matrix and can generally show exceptional points. However, whether such EPs can be reached with a physical set of parameters is entirely dependent on the configuration of the system and baths. Since Eq. \eqref{eq:eomm} in general can produce non-Markovian reduced dynamics of the system modes, the evolution matrix $A$ represents a suitable pathway to investigate non-Markovian EPs. Let us note that this approach is equivalent to one involving Green's functions.  For fermions, the retarded Green's function of the system--RCs is given by $G^R(t) = -i \Theta(t) \expval{ \{\vec v(t), \vec v^\dagger(0)\}} = - i \Theta(t) D(t)$ (for bosons, the anticommutator is replaced by a commutator) . In frequency space, it is the resolvent $ G^R(\omega) = (\omega \mathds 1 - iA)^{-1}$. Therefore, EPs correspond to $G^R(\omega)$ having higher-order poles, i.e., a Jordan block of size $m>1$ in $A$ produces a pole of order $m$ in the resolvent. 

Finally, we note that Lorentzian baths have been considered in the context of non-Markovian EPs in spin-boson models using pseudomode equations of motion (PMEOM)\cite{Lin2025,Zhang20261}. Heisenberg-picture formulations have also been used to investigate non-Markovian EPs, including the adjoint PMEOM for linear bosonic systems \cite{Lin2025} and Heisenberg--Langevin approaches \cite{Shen_2026,Ghosh2026}. The key advantage of using Heisenberg equations is the significant reduction in the dimension of the involved evolution operators. For example, one fermionic dot and one fermionic RC have a four-dimensional Hilbert space, corresponding to a $16\times16$ full Liouvillian representation, whereas their linear Heisenberg equations involve only a $2\times2$ matrix $A$, as we will see later. For $m=N_s+N_b$ fermionic modes, there is a $4^m\times4^m$ Liouvillian, whereas the closed linear Heisenberg equations involve an $m\times m$ evolution matrix. This reduction exploits the quadratic, number-conserving structure and facilitates the calculation of retarded dynamics and its EPs. There is further a difference in the order of any exceptional points, which is explained below.

\subsection{Dynamics and the order of the EP}

Suppose $A$ has a Jordan block of size $N$ at eigenvalue $\lambda$,
while its remaining eigenvalues are non-defective. Eq. \eqref{eq:eomsol} can then be expressed in the form \cite{Khandelwal2021} 
\begin{equation}
    \begin{aligned}
    \vec v(t) = &e^{\lambda t} \sum_{l=1}^N\left(\sum_{m=0}^{l-1}\frac{b_{l-m}}{m!}t^m\right)\vec \rho^{\hspace*{0.1cm}\prime}_l + \sum_{l=N+1}^{N_s+N_b}c_l e^{\lambda_l t}\vec \rho_l\\
    &+ \int_0^t ds D(t-s)\vec \xi(s).
\end{aligned}
\end{equation}with $b_k,c_k\in \mathbb C$, $\vec\rho_k$ the eigenvectors of $A$ and $\vec \rho_k^{\hspace*{0.1cm}\prime}$ the generalised eigenvectors. As expected, this expression contains polynomial factors of degree at most ${N-1}$. However, this property only pertains to the evolution of system and reaction-coordinate operators, and not to arbitrary physical observables (Hermitian products of system and RC operators). Any such observable that is constructed as a sum of $p$-product of the system and RC operators will have such polynomial factors of degree at most ${p(N-1)}$, reflecting a higher-order EP behaviour. 
Let us mention how this observation leads to a key difference between the EPs of the Heisenberg evolution matrix and those of the corresponding PMEOM Liouvillian. The PMEOM describes the evolution of the joint system--RC density matrix, which can be constructed through products of system--RC operators. As a result, the order of the EP in the PMEOM Liouvillian can be larger compared to Heisenberg equations, even if the underlying physics remains the same. The exact relation between EP orders depends on the considered system, and will be discussed in examples below. The Heisenberg formulation directly exposes the EP structure of
the linear operator dynamics and provides a starting point for determining its signatures in higher-order observables.

\subsection{Signature of non-Markovianity}
 The paradigmatic approach to quantify non-Markovianity is through information backflow. This is captured by the distinguishability, quantified by the trace distance \cite{Breuer2009},
\begin{align}
    \mathcal D(t,\rho_1,\rho_2) = \frac{1}{2}||\rho_1-\rho_2 ||_1
\end{align} where the two states evolve under the same dynamical map. CP-divisible dynamics cannot increase the trace
distance between two evolving states. Intervals
with $\sigma(t)>0$ therefore witness information
backflow, a signature of non-Markovianity. The distinguishability can be used to construct the Breuer-Laine-Piilo (BLP) measure for non-Markovianity, 
\begin{align}
    \label{eq:blp}
    \mathcal N_{\text{BLP}} = \max_{\rho_1(0),\rho_2(0)}\int_{\sigma>0}\sigma(t) dt
\end{align} with $\sigma(t)= \frac{d\mathcal D(t,\rho_1,\rho_2)}{dt}$. A non-zero $\mathcal N_{\text{BLP}}$ signals information backflow, which is a consequence of non-Markovian evolution. We define a revival as an interval during which the distinguishability increases, bounded by periods of decreasing distinguishability.

\section{Model}
\label{sec:mod}

The main model we consider in this work is that of a chain of $N_s$ quantum dots coupled with Lorentzian baths. We consider the system Hamiltonian
\begin{align}
    H_{\text{S}} = \sum_{k=1}^{N_s} \epsilon_d \hat d_k^\dagger \hat d_k + \sum_{\substack{ k=1}}^{N_s-1}g(\hat d^\dagger_k \hat d_{k+1}+\hat d^\dagger_{k+1} \hat d_{k}). 
\end{align}We assume that all dots have the same energy $\epsilon_d$ and that each Lorentzian bath spectral density is centred at this energy. The corresponding reaction coordinates therefore also have energy $\epsilon_d$. For simplicity, we assume that all inter-dot interactions are equal. Overall, we operate within the scenario discussed in Sec. \ref{sec:gen}, where the system--bath interaction takes exactly the form given in Eq. \eqref{eq:mod}, with $\hat S_k = \hat d_k$. Following the procedure described above, the evolution of the dot--RC setup can be expressed as a Langevin equation with the non-Hermitian evolution matrix $A$. We let the characteristic polynomial of $A$ be $ P(\lambda) = \det(\lambda \mathds 1 - A)$.

\subsection{Symmetry in the eigenvalues}

 First, note that the eigenvalues of $A$ can be expressed as $\lambda_j = -i\epsilon_d+\tilde\lambda_j$, where $\tilde \lambda_j$ are the eigenvalues of the shifted evolution matrix $\tilde A \coloneqq A+i\epsilon_d\mathds 1$. For the chain configuration, we show in App. \ref{app:chain} that there exists an invertible matrix $S$ such that $\tilde A'=S\tilde AS^{-1}$ is a real matrix and therefore has a characteristic polynomial with real coefficients. Since $\tilde A$ and $\tilde A'$ are related by a similarity transformation, they have the same characteristic polynomial with real coefficients. This property leads to a symmetric structure in the eigenvalues. First, the eigenvalues of $\tilde A$ must either be real or come in complex conjugate pairs. A similar property of eigenvalues holds for quantum dynamical semigroups \cite{alicki2007}, which leads to a symmetric structure of Liouvillian exceptional points \cite{Hatano2019,Khandelwal2021}.  Second, the discriminant $\Delta$ of the characteristic polynomial determines the parity of the number of complex conjugate pairs. Here we apply the root classification to the shifted characteristic polynomial, whose discriminant is unchanged by the common eigenvalue shift. For a characteristic polynomial $ P(\lambda) = a\prod_{i} (\lambda-\lambda_i)$ with roots $\lambda_i$, the discriminant takes the form $\Delta = a^{2n-2} \prod_{i<j}(\lambda_i-\lambda_j)^{2}$. For $\Delta>0$, the roots are distinct and there is an even number of complex-conjugate pairs. For $\Delta<0$, there is at least one pair of complex-conjugate roots. Lastly, $\Delta=0$ implies at least one repeated root. An EP of order $k$ requires an eigenvalue of algebraic multiplicity at least $k$. The repeated-root criterion also holds for polynomials with complex coefficients, whereas the preceding sign classification requires real coefficients. For smaller dimensions of $A$, these will lead to further simplifications, as we will see in later sections.

\section{Case A: second-order EP}
\label{sec:caseA}
We consider a single-level quantum dot coupled with one Lorentzian bath. The system can be exactly described by adding one RC into the picture. Following the method described above, the joint evolution of the system and RC is given by
\begin{align}\label{eq:eom1}
   \dot{\vec v} = A\vec v + \vec \xi,\,\,\,\, \text{with}\,\,\,\, A = -\begin{pmatrix}
        i\epsilon_d &i\alpha\\
        i\alpha &i\epsilon_d + \delta
    \end{pmatrix},
\end{align}with $\vec v=(\hat d,\hat a)^T$. $\vec \xi = (0,\xi)^T$, with $\xi = -i \sum_\alpha T_\alpha e^{-i\omega_\alpha t}\hat b_\alpha(0)$ assuming initial time $t=0$. The Heisenberg evolution operator $A$ has eigenvalues $\lambda_{1,2} = -\frac{1}{2}(2i\epsilon_d + \delta \pm \eta)$, with $\eta = \sqrt{\delta^2-4\alpha^2} = \Delta_{\text{A}}^{1/2}$. The discriminant $\Delta_{\text{A}}=0$ or $\delta=2\alpha$ (with $\delta>0$) marks a second-order EP or EP2. Moreover, this eigenvalue structure has an analogy with the damped harmonic oscillator, with $\Delta_{\text{A}}>0$ corresponding to overdamping, $\Delta_{\text{A}}<0$ to underdamping and $\Delta_{\text{A}}=0$ to critical damping \cite{Khandelwal2021,Khandelwal2025em}. 

For states diagonal in the occupation basis, the state of the quantum dot is completely characterised by its average population $n_d(t)= \langle\hat d^\dagger(t)\hat d(t)\rangle$. Assuming no initial system--RC--bath correlations, Eq. \eqref{eq:eom1} can be solved to give 
\begin{equation}\label{eq:eomsol1}
    \begin{aligned}
&n_d(t) =   |D_{11}(t)|^2n_d(0) + |D_{12}(t)|^2 n_a(0) \\
&+ \delta\hspace{-0.1cm}\int \frac{d\omega}{\pi} \int_0^t \hspace*{-0.1cm}ds \int_0^t \hspace*{-0.1cm}ds' D^*_{12}(t-s)D_{12}(t-s')e^{-i\omega(s'-s)}f(\omega),
\end{aligned}
\end{equation}where $D(t)\coloneqq e^{At}$, $n_a(0)=\langle\hat a^\dagger(0)\hat a(0)\rangle$, and $f(\omega)$ is the Fermi--Dirac distribution of the residual bath. Here we choose an initially factorised
dot--RC--residual-bath state and set
$n_a(0)=f(\epsilon_d)$.

\subsection{Onset of non-Markovianity}

We use the distinguishability as a signature of non-Markovianity. In the case of a single-level quantum dot, assuming no initial coherence, the density matrix is diagonal at all times. The trace distance simply reduces to the difference between two average populations, $\mathcal D(t,n_d^{(1)},n_d^{(2)}) = |n_d^{(1)}(t)-n_d^{(2)}(t)|$, where the two states evolve from different initial populations. Eq. \eqref{eq:eomsol1} yields $\mathcal D(t,n_d^{(1)},n_d^{(2)}) = D^*_{11}(t)D_{11}(t)|n_d^{(1)}(0)-n_d^{(2)}(0)|$. Using the analytical form of $D(t)$, we obtain for the three cases,
\begin{equation}\label{eq:rate}\begin{aligned}
 &D^*_{11}(t)D_{11}(t)   = \\&\begin{cases}
            \frac{1}{4}e^{-\delta t}(2+\delta\, t)^2, & \hspace*{-0.2cm} \text{EP}\\
			e^{-\delta t}\big(\cosh(\eta t/2)+\frac{\delta}{\eta}\sinh(\eta t/2)\big)^2 , & \hspace*{-0.2cm} \text{overdamped} \\
            e^{-\delta t}\big(\cos(\eta' t/2)+\frac{\delta}{\eta'}\sin(\eta' t/2)\big)^2 , &  \hspace*{-0.3cm} \text{underdamped}
		 \end{cases}
 \end{aligned}
\end{equation}with $\eta'=\sqrt{4\alpha^2-\delta^2}>0$. We therefore find that at the EP and in the overdamped situations, there is monotonic decline in the distinguishability, with revivals impossible. However, the underdamped regime $\delta^2<4\alpha^2$ shows oscillatory behaviour, i.e., revivals of distinguishability. In this way, the EP marks the onset of non-Markovian revivals. This property has been noted in \cite{Lin2025} by investigating the decoherence function associated with a two level system in a spin-boson model. Let us also contrast this with the wideband limit ($J(\omega)\equiv \Gamma$), in which there is no EP. Here, it can be shown that $\mathcal D(t,n_d^{(1)},n_d^{(2)})=e^{-\Gamma t}|n^{(1)}_d(0)-n^{(2)}_d(0)|$, which is purely exponentially decaying. As expected, in the wideband limit, there is no revival of quantum information.

 We restrict the initial single-dot states to those
diagonal in the occupation basis. Due to the form of the distinguishability in this case, the BLP  measure can be calculated analytically. Maximisation over these initial states is achieved by setting $|n_d^{(1)}(0)-n_d^{(2)}(0)|=1$. The BLP measure can then be shown to be exactly (see App. \ref{app:blp}),
\begin{align}\label{eq:nblp}
    \mathcal N_{\text{BLP}} = \frac{1}{e^{\frac{2\pi \delta}{\eta'}}-1},
\end{align}
if the setting is underdamped, and zero otherwise. Naturally, the highly non-Markovian limit corresponds to the Lorentzian spectral density being sharply peaked ($\delta\ll\alpha$), i.e. when the square root factor $\eta' = \sqrt{4\alpha^2-\delta^2}$ becomes very large compared to $\delta$. Strictly, $\alpha/\delta\to\infty$ (equivalently, $\eta'/\delta\to\infty$) leads to $\mathcal N_{\text{BLP}}\to\infty$. On the other hand, getting closer to the EP, $\delta\to2\alpha^-$,  $\delta/\eta'\to\infty$ and $\mathcal N_{\text{BLP}}\to0$. Therefore, not only does the non-Markovian EP demarcate the onset of information backflow, the structure of the involved eigenvalues (through the discriminant $\Delta_{\text{A}}$) fundamentally determines the level of non-Markovianity in the evolution. 

Alternatively, one may work with exact master equations \cite{Breuer2007}. For an initially empty RC and residual bath, it can be shown that the damping rate becomes negative (i.e., the evolution becomes CP-indivisible) only in the case of underdamping, and is non-negative otherwise.  Let us note that this property has been observed long ago \cite{Lu2010} with signatures of non-Markovianity, without noting an underlying connection with exceptional points.

\section{Case B: third-order EP}
\label{sec:caseB}

 We now consider a single-level quantum dot coupled with two baths with Lorentzian spectral densities. The situation can be exactly described by adding two RCs into the picture, each of which interacts with a residual bath with a flat spectral density. As in Eq. \eqref{eq:eom1}, this joint evolution is described by $\dot{\vec v} = A\vec v + \vec \xi$, with $\vec v=(\hat d,\hat a_1,\hat a_2)^T$, $\vec \xi=(0,\xi_1,\xi_2)^T$ (where $\xi_i = -i \sum_\beta T_{i\beta} e^{-i\omega_{i\beta} t}\hat b_{i\beta}(0)$) and 
\begin{align}\label{eq:eom2} A = -\begin{pmatrix}
          i\epsilon_d &i\alpha_1&i\alpha_2\\
         i\alpha_1 &i\epsilon_d+\delta_1  &0\\
        i\alpha_2 &0  &i\epsilon_d + \delta_2
    \end{pmatrix}.
\end{align}For equivalent spectral densities ($\delta_1=\delta_2\equiv \delta$ and $\alpha_1=\alpha_2\equiv\alpha$), $A$ has a reducible characteristic polynomial and two of the eigenvalues have a quadratic structure determined by the discriminant $ \delta^2-8\alpha^2$. Naturally, this factor  determines the onset of oscillations at the corresponding EP $\delta_{\text{EP}}=2\sqrt{2}\alpha$. The factor in Eq.~\eqref{eq:rate}, as well as the measure $\mathcal N_{\text{BLP}}$ take an identical form, except with a redefined discriminant. Compared to the case with a single Lorentzian bath, the only change is in the position of the EP, which has shifted to a larger value of $\delta$. This shift follows from the increased effective dot--RC coupling
$\alpha_{\mathrm{eff}}=\sqrt{\alpha_1^2+\alpha_2^2}=\sqrt{2}\alpha$,
which allows revivals to persist at larger spectral widths.

\begin{figure}
    \centering
    \includegraphics[width=0.5\textwidth]{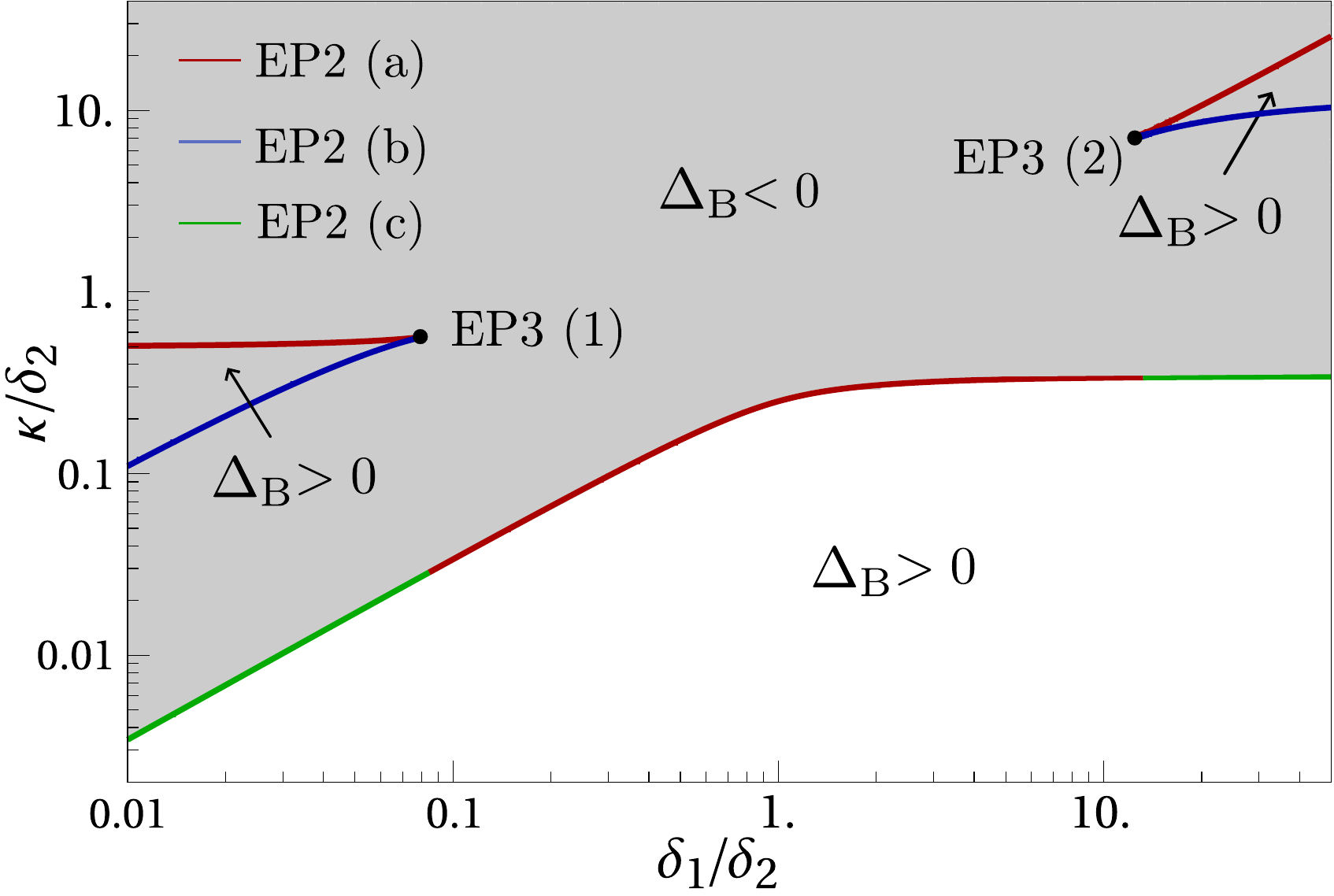}
    \caption{EP lines for Eq.~\eqref{eq:eom2} with $\kappa_1=\kappa_2=\kappa$ and regions showing a non-zero BLP measure defined in Eq.~\eqref{eq:blp} in grey. Note that with a characteristic polynomial of degree 3, a negative discriminant is now a \textit{sufficient} but not \textit{necessary} condition for non-Markovian revivals.}
    \label{fig:ep3}
\end{figure}

The situation is more complicated for a general set of parameters and the eigenvalues of $A$ come as cubic roots. For simplicity we set the couplings to baths equal, $\kappa_1=\kappa_2\equiv \kappa$. Furthermore, we focus on the shifted evolution matrix $\tilde A\coloneqq A+i\epsilon_d \mathds 1$, with eigenvalues $\tilde \lambda_i$. Making use of Cardano's formula, the eigenvalues of $\tilde A$ take the following general form,
\begin{equation}
    \begin{aligned}
        &\tilde \lambda_1 = a+ u + v\\
        &\tilde \lambda_2 = a + \omega u +\omega^2v\\
        &\tilde \lambda_3 = a + \omega^2 u + \omega v,
    \end{aligned}
    \label{eq:roots}
\end{equation}where $a =  -(\delta_1+\delta_2)/3$, $\omega = e^{2\pi i/3}$, $u=\sqrt[3]{q+\sqrt{-\Delta_{\mathrm B}/108}}$, $v=\sqrt[3]{q-\sqrt{-\Delta_{\mathrm B}/108}}$, with the discriminant $\Delta_{\mathrm B}=-108(q^2+p^3)$,
and cube-root branches chosen such that $uv=-p$.  The explicit forms of $p$ and $q$ as functions of the parameters are given in App. \ref{app:ep3}. For $\Delta_{\mathrm B}>0$, all three eigenvalues of $\tilde A$
are real and distinct. For $\Delta_{\mathrm B}<0$, one eigenvalue
is real and the other two form a complex-conjugate pair. Eigenvalues with this structure, in principle, permit two types of EPs \cite{Hatano2019}:
\begin{enumerate}
\item{A second-order EP (EP2) at $\Delta_{\text{B}}=0$ with $(p,q)\neq(0,0)$.}
\item{A third-order EP (EP3) at $p=q=0$.}
\end{enumerate}In general, whether such EPs can be reached for a physical set of parameters is highly dependent on the setting. Here, both types of EPs are reachable; we briefly explain this here and give further details in App. \ref{app:ep3}. The condition $\Delta_{\mathrm B}=0$ is a cubic equation in
$\kappa$, whose three algebraic solution branches
$\kappa_{\mathrm{EP}}^{(j)}(\delta_1,\delta_2)$, $j=1,2,3$,
are given in App.~\ref{app:ep3}. Their real, positive portions
define the exceptional-point surfaces labelled (a), (b), and (c).
These correspond to EP2s except where $p=q=0$, at which the
degeneracy becomes an EP3. In particular, the branches labelled
(a) and (b) meet along the two EP3 lines labelled (1) and (2),
determined by
\begin{equation}
    \begin{aligned}
    \frac{\delta_1}{\delta_2}
    &=\frac{1}{2}\left(
    7+3\sqrt{3}\pm\sqrt{42\sqrt{3}+72}
    \right),\\
    \frac{\kappa}{\delta_2}
    &=2+\sqrt{3}\pm\frac{1}{3}\sqrt{26\sqrt{3}+45},
    \end{aligned}
    \label{eq:ep3cond}
\end{equation}
where the signs must be chosen consistently. Fig.~\ref{fig:ep3} shows a cross section at $\delta_2=1$,
in which the EP2 surfaces become curves and the EP3 lines
become points. The two EP3 lines are mapped into one another
under the exchange $\delta_1\leftrightarrow\delta_2$.

\subsection{Onset of non-Markovianity}
The discriminant $\Delta_{\text{B}}$ distinguishes two spectral regimes of the shifted evolution matrix $\tilde A$:
for $\Delta_{\text{B}}>0$, all three eigenvalues are real, whereas for $\Delta_{\text{B}}<0$, there is one real eigenvalue and a complex-conjugate pair. For the resonant single dot coupled to two Lorentzian baths considered here, $\Delta_{\text{B}}<0$ is sufficient, but not necessary, for information backflow (see Fig.~\ref{fig:ep3}). A demonstration of sufficiency is found in App. \ref{app:ep3}. For $\Delta_B > 0$, the eigenvalues of the shifted evolution matrix $\tilde A$ are all real, while for $\Delta_B < 0$, there is a complex conjugate pair. A complex-conjugate pair of eigenvalues naturally leads to oscillations in the decay of the distinguishability $\mathcal D(t)$. Therefore, $\Delta_{\text{B}}<0$ demarcates a non-Markovian regime. We note that in this case, the positivity of $\sigma(t)$ is determined by a transcendental equation and therefore $\mathcal N_{\text{BLP}}$ has been calculated numerically. For $\Delta_{\text{B}}>0$, all three eigenvalues are real and there are no oscillatory eigenmodes. However, this does not necessarily mean that the distinguishability decays monotonically. We show in App. \ref{app:ep3} that whenever the shifted eigenvalue $\tilde\lambda$ corresponding to the slowest decaying mode fulfils
\begin{equation}
    \abs{\tilde \lambda} > \frac{2\delta_1\delta_2}{\left(\delta_1+\delta_2\right)},
\end{equation}
the distinguishability is not monotonically decaying.

\section{Case C: quartic structure and EP4}
\label{sec:caseC}
To investigate EP4, the natural path forward would be to consider the spectral density to be a sum of three Lorentzian functions. There are now three reaction coordinates and $A$ is $4\times4$. Let us first note that the corresponding spectral density leads to a situation where an EP4 cannot be reached for real and positive parameters. In fact, for a single quantum dot coupled to an arbitrary number of Lorentzian baths, EPs of order greater than 3 are impossible (see App. \ref{app:ep4pr}). There are essentially two natural ways around this to investigate a quartic structure of the eigenvalues while being able to reach an EP4.

\subsection{Band gap structure}
Consider a single quantum dot dissipating through the following band-gapped spectral density instead
\begin{align}\label{eq:Jband}
    J_{\text{\tiny{BG}}}(\omega) = \sum_{i=1}^2\frac{\kappa_i\delta_i^2}{(\omega-\epsilon_d)^2+\delta_i^2} -  \frac{\kappa_2\delta'^2}{(\omega-\epsilon_d)^2+\delta'^2}.
\end{align}
This corresponds to two non-interacting baths: one Lorentzian bath and one Lorentzian bath with a band-gap structure (see, for example \cite{Breuer2007}). Band-gap structures have been explored in the context of non-Markovian EP3 \cite{Lin2025}. Interestingly, this structure is not necessary to reach an EP3, as we have shown in Sec. \ref{sec:caseB}, but makes it possible to reach an EP4. The retarded dynamics associated with the spectral density in Eq.~\eqref{eq:Jband} admits the following auxiliary matrix representation,
\begin{align}
    A = -\begin{pmatrix}
        i\epsilon_d & i\alpha_1 & i \alpha_2 & \alpha'\\
        i\alpha_1 &i\epsilon_d+\delta_1 &0 &0\\
        i\alpha_2 &0 &i\epsilon_d+\delta_2 &0\\
        \alpha' &0 &0 &i\epsilon_d+\delta'
    \end{pmatrix},
    \label{eq:bandgap}
\end{align}where $\alpha' = \sqrt{\kappa_2\delta'/2}$ and $\alpha_{1,2}$ are as defined previously. Here, the negative Lorentzian contribution is represented by an auxiliary variable rather than a physical reaction-coordinate mode; the resulting equations nevertheless reproduce the exact retarded dot dynamics.

To explore EPs in this model, we first make two simplifications; although a general case can be considered, our approach leads to neat expressions, while capturing all the necessary properties. Let us first fix $\delta' =\delta_2/2$. Next, to simplify the eigenvalues, we turn the characteristic polynomial of $A$ into a  biquadratic form by choosing $\kappa_1 = (4\delta_1^2-\delta_2^2 + 4\delta_2\kappa_2)/(8\delta_1)$. Physical parameters additionally require $4\delta_1^2-\delta_2^2+4\delta_2\kappa_2>0$, so that $\kappa_1>0$. The eigenvalues then take the form $\alpha \pm \sqrt{\beta\pm\sqrt{\gamma}}$. The expression of $\alpha$, $\beta$ and $\gamma$ can be found in App. \ref{app:bandgap}. The condition $\gamma=0$, with $\beta\neq0$, gives
two double eigenvalues. Alternatively, a double
eigenvalue occurs when $\beta^2=\gamma$ with
$\beta\neq0$. A quadruple eigenvalue requires
$\beta=\gamma=0$. It can be checked that these conditions correspond to EP2 and EP4, respectively.

The EP2 condition $\gamma=0$ translates to
\begin{align}
   \text{EP2}:\,\, \frac{\kappa_2}{\delta_2} = -\frac{3}{2}\frac{\delta_1}{\delta_2} + \frac{5}{4} \pm \sqrt{2\frac{\delta_1^2}{\delta_2^2}+1-3\frac{\delta_1}{\delta_2}},
   \label{eq:bgEP2}
\end{align}which corresponds to two curves in the space of $\{\kappa_2,\delta_1\}$. The EP4 is naturally located precisely where these two curves meet, i.e., when
\begin{align}
\text{EP4}:\,\, \frac{\kappa_2}{\delta_2} = \frac{\delta_1}{\delta_2} = \frac{1}{2}.
\label{eq:bgEP4}
\end{align}Although $A$ exhibits an EP4, its full Jordan structure
is not visible in the dot propagator at this parameter
point. The cancellation between the spectral contributions reduces the local dot Green function to that of a single Lorentzian bath at critical damping. Next, we look at a model that retains the EP4 in its dynamics.

\begin{figure}
    \centering
\includegraphics[width=1\linewidth]{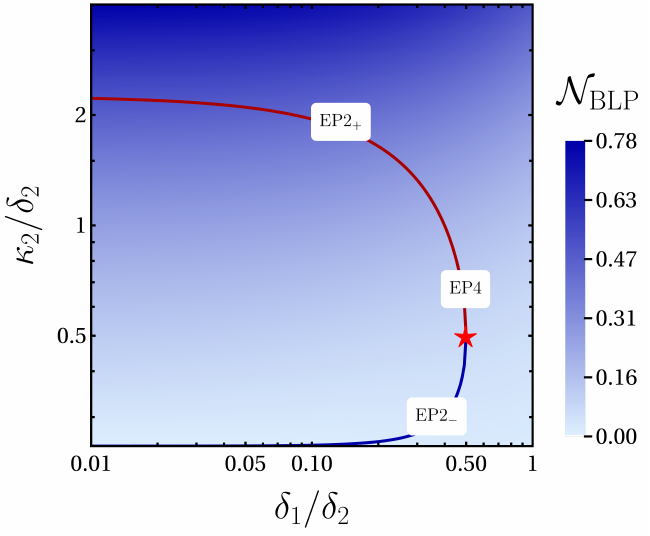}
    \caption{BLP measure for dynamics given by Eq.~\eqref{eq:bandgap}. Red and blue curves show EP2$\pm$ branches of Eq.~\eqref{eq:bgEP2} with EP4 determined by Eq.~\eqref{eq:bgEP4} highlighted. Note that the EP lines no longer demarcate the boundary between Markovian and Non-Markovian dynamics.}
    \label{fig:bgep4BLP}
\end{figure}

\subsection{Coupled dots}
\begin{figure}
    \centering
    \includegraphics[width=\linewidth]{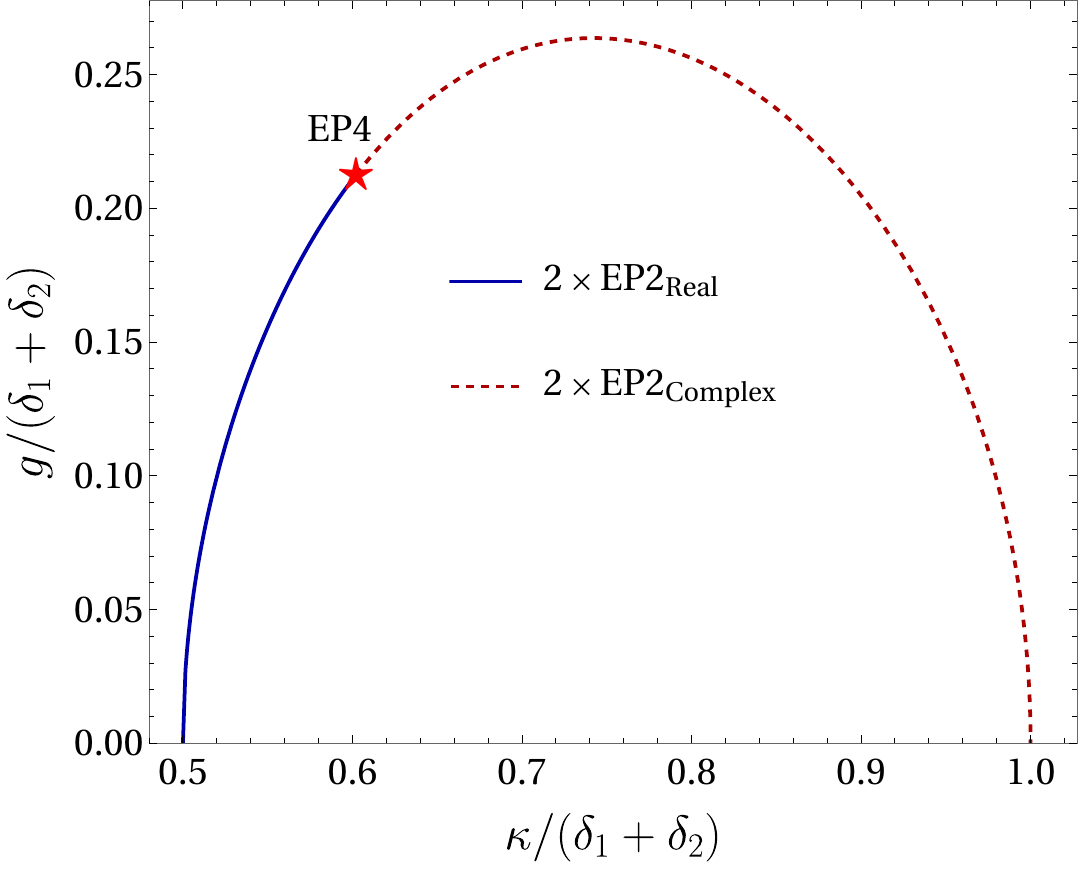}
    \caption{Parametric sweep of Eq.~\eqref{eq:ep4cond} with respect to $q$ for $\delta_1+\delta_2 = 1$ and $q \in [0,1/4]$ plotted as a function of corresponding $g,\kappa$ values. As $q$ approaches $q_\star$ from 0 (blue curve) the two double eigenvalues of $\tilde A$ are real. Beyond the EP4 the real part of the eigenvalues is constant ($-S/4$) with an increasing imaginary component.}
    \label{fig:paramsweepEP4}
\end{figure}

We now consider two coupled quantum dots, each interacting with a bath. The evolution of $\vec v=(\hat d_1,\hat d_2,\hat a_1,\hat a_2)^T$ is governed by $\dot{\vec v} = A\vec v + \vec \xi$, with
\begin{align}
    A = -\begin{pmatrix}
        i\epsilon_d & ig & i \alpha_1 & 0\\
        ig &i\epsilon_d &0 &i\alpha_2\\
        i\alpha_1 &0 &i\epsilon_d+\delta_1 &0\\
        0 &i\alpha_2 &0 &i\epsilon_d+\delta_2
    \end{pmatrix}
    \label{eq:coupled}
 \end{align}and $\vec \xi = (0,0,\hat \xi_1,\hat \xi_2)^T$ constructed as in previous examples.
For simplicity, we consider $\kappa_1=\kappa_2=\kappa$. We define $S= \delta_1+\delta_2$ and $P = \delta_1\delta_2$. At the EP4 all eigenvalues are equal to $-S/4$. We look for solutions of the characteristic polynomial of $\tilde A$ such that it takes the form $\tilde P(\lambda) = ((\lambda+S/4)^2-r^2)^2 = \left(\lambda+S/4-r\right)^2\left(\lambda+S/4+r\right)^2 $, describing two simultaneous EP2s for $r\neq0$ that coalesce into an EP4 at $r=0$. Expanding the characteristic polynomial of $\tilde A$ and matching terms with this chosen form, it is possible to reduce the equations into a quadratic equation for $r^2$ only in terms of the ratio $q=P/S^2$. The solution of this quadratic corresponding to physical values of $g,\kappa$ is given by
\begin{equation}
    \frac{r^2}{S^2} = \frac{16q^2-13q+4\sqrt{q}(2q-1)+1}{16(1-q)},
    \label{eq:ep4cond}
\end{equation}
where
\begin{align}
    &\kappa = \frac{1}{2(1-\sqrt{q})}S, \hspace{6pt}
    g = \sqrt{\frac{(1-4q)\sqrt{q}}{4(1-\sqrt{q})}}S
\end{align}
and $0 < q < 1/4$ such that $g$ remains real. This interval corresponds to positive, unequal bath widths $\delta_1,\delta_2$. In terms of $S,q$ we have
\begin{equation}
    \delta_{1/2} = \frac{S}{2}\left(1\pm\sqrt{1-4q}\right),
\end{equation}
Note that the parameter point $\delta=(\delta_1,\delta_2)$ is equivalent to one with swapped $\delta_1,\delta_2$. The EP4 then corresponds to the solution of $r^2=0$, which has only one valid real solution when $0 \leq q \leq 1/4$ given by $q_{\star} \approx 0.028761$ (See Fig. \ref{fig:paramsweepEP4}).  

\begin{figure}
    \centering
    \includegraphics[width=\linewidth]{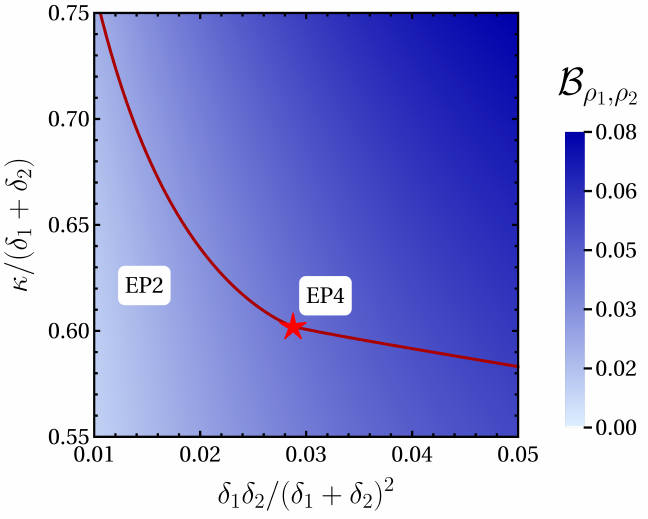}
    \caption{Integrated information backflow $\mathcal{B}_{\rho_1,\rho_2}=\int_{\sigma>0}\sigma(t)\,dt$ for the specified initial pair, for dynamics given by Eq.~\eqref{eq:coupled}. The red solid curve corresponds to second-order EPs determined by $\Delta=0$, with the EP4 highlighted. EPs no longer demarcate regions of revival/no-revival of distinguishability for the specified initial pair of states. Parameter choices: State 1 - $n_1(0) = 1, n_2(0) = 0$, State 2 - $n_1(0) = n_2(0) = 0$, both states have identical environmental preparations and initially empty RCs, $S=1, g/S=g_{\text{EP4}}/S\approx0.21$.}
    \label{fig:ep4Surf}
\end{figure}

\subsection{Onset of non-Markovianity}
We have found in the cases of quadratic eigenvalues of Model A that the EP2 curves mark the transition between Markovian and non-Markovian regimes. We then showed that this correspondence weakens in case B. This property is fundamentally rooted in the fact that for these cases, the discriminant $\Delta$ solely determines whether the eigenvalues are real ($\Delta>0$) or come as complex conjugate pairs ($\Delta<0$). In Cases A and B, a complex-conjugate pair of shifted eigenvalues is sufficient for information backflow. Since the EPs are determined by setting $\Delta=0$, it is therefore natural that they would demarcate this transition. However, the situation is fundamentally different for quartic eigenvalues. Here, the discriminant alone is insufficient to determine whether the eigenvalues are real or complex. A quartic discriminant does not distinguish four
real eigenvalues from two complex-conjugate pairs. EP curves can therefore separate regions that already contain oscillatory modes on both sides. Their relation to information backflow must be determined from the reduced dynamics.

Fig. \ref{fig:ep4Surf} shows a calculation of the information backflow with the maximally mixed state as initial state for the two dots in a region of parameter space for a fixed $g_\star/S$ corresponding to the value of $g$ at the EP4 with the contour $\Delta=0$ overlaid. Note that in the parameter region shown, the EP curve does not separate regions with and without information backflow.

%\FloatBarrier
\section{Dynamical consequences of non-Markovian EPs}
\subsection{Information revival}
\label{sec:inf}

The preceding examples show that exceptional points need not mark the boundary between monotonic decay and information backflow. We now investigate information backflow specifically \emph{at} EPs and ask how many distinct revivals of distinguishability are possible. 
%\par\smallskip\noindent\gb{GB -- Could we add: ``We ask how many separate recoveries of local distinguishability are possible at an EP, rather than whether the EP marks their onset''? EP3 in case B and EP4 in coupled case C are the highest accessible orders in these respective configurations. Is there a reason why we focus on the highest order EP only?}\par\smallskip

We consider the setup of Sec. \ref{sec:gen} with several quantum dots coupled to non-interacting Lorentzian baths. To construct the distinguishability, we consider two initial states that differ only in the occupation of one observed dot $j$: $n_j^{(1)}(0)=1$ and $n_j^{(2)}(0)=0$. We assume that all other dots have the same initial preparation and that the RCs and residual baths have the same thermal preparation. The initial state is factorised between the dots, RCs, and residual baths; there are no initial coherences. Here $\rho_1$ and $\rho_2$ refer to the two reduced states of the selected $j$-th dot. It is simple to see that the exact solution gives
\begin{equation}\label{ir:contrast}
\mathcal D^{(j)}(t,\rho_1,\rho_2)=|n_j^{(1)}(t)-n_j^{(2)}(t)|=|D_{jj}(t)|^2.
\end{equation}Contributions from the common initial occupations and from the residual noise cancel in the population difference. This approach involving the reduced state of a selected dot allows us to probe information backflow into that dot, with the remaining dots, RCs, and baths constituting its environment.

Consider case B, i.e., a single dot coupled with two Lorentzian baths. In this model, every EP3 produces exactly one positive-time zero of the distinguishability and exactly one revival. At the EP3, the distinguishability decreases to zero, revives once, and then decays to zero at long times. App. ~\ref{app:71} proves that no additional revivals occur. We emphasise that it can be shown that this result holds for any number of independent Lorentzian baths centred at \(\epsilon_d\).

Now consider Case C, with coupled quantum dots, each interacting with its own Lorentzian bath. For generality, we relax the assumption of equal couplings $\kappa_{1,2}$. This situation leads to a richer structure and different dots can have different revival counts at the same EP. We assume $\delta_1>\delta_2>0$, and perform the preparation above separately for each observed dot. Table \ref{tab:1} gives the number of revivals at any physical EP4 in this model, dividing the parameter space into two regimes.

\begin{table}[h!]
\begin{center}
\begin{tabular}{ |c|c|c| }
\hline
Width ratio & Dot 1 & Dot 2\\
\hline
$1<\frac{\delta_1}{\delta_2}<3$ & One revival & Monotonic decay\\
$\frac{\delta_1}{\delta_2}>3$ & Two revivals & One revival\\
\hline
\end{tabular}\caption{Number of revivals at any EP4 in the coupled-dot model (case C). No physical EP4 exists for $\delta_1/\delta_2=1,3$. The EP4 discussed in Sec. \ref{sec:caseC} belongs to the second row.}\label{tab:1}
\end{center}
\end{table}
 Proofs of the given numbers are presented in App.~\ref{app:72}. Although both dots share the same exponential decay at the EP4, their distinguishability can exhibit different numbers of revivals.

Let us now consider a more general scenario with any geometry of $N_s$ coherently coupled fermionic quantum dots, each interacting with one or more independent Lorentzian baths centred at the common dot energy $\epsilon_d$. The RC mapping introduces one fermionic RC per bath, giving an enlarged evolution matrix $A$ of dimension $m=N_s+N_b$.  When all $m$ eigenvalues and eigenvectors coalesce into a single EP of order $m$,
\begin{equation}\label{ir:bound}
\begin{aligned}
\text{number of revivals of }\mathcal D^{(j)}(t,\rho_1,\rho_2)&\leq m-2.
\end{aligned}
\end{equation}Therefore, the number of revivals is limited by the maximal order of the EP. The proof can be found in App. \ref{app:73}. Let us note, however, that this maximal-order EP may not be physically reachable in some models, as we have seen earlier.

\subsection{Relaxation rates}
Above we have investigated the role of EPs to predict the revival of information. Below we complement this discussion by investigating how the rate of loss of information of an initial state relates to EPs.

Consider
\begin{equation}
    R(t) = \frac{\mathcal D(t,\rho_1,\rho_2)}{\mathcal D(0,\rho_1,\rho_2)},
\end{equation}
the normalized distinguishability between two evolving initial states with identical bath preparation.
We can then define \cite{Mouloudakis2022},
\begin{equation}
    \Gamma(t) = -\frac{\log R(t)}{t},
\end{equation} 
$\Gamma$ then corresponds to a time-averaged logarithmic rate of information loss. 
EPs can enhance asymptotic relaxation under suitable parameter constraints \cite{Khandelwal2021,zhang2022,Khandelwal2025em,Zhang20261}. At long times, $\Gamma$ is determined by the eigenvalue $\mu$ of $A$ with largest (least negative) real part. The corresponding decay rate is bounded by the trace of $A$ \cite{Kumar2022a},
 \begin{equation}
     \abs{\Re (\mu)} \leq  \frac{1}{m}\sum_i \abs{\Re (\lambda_i)}= \frac{\abs{\Re( \tr(A))}}{m}, 
 \end{equation}
as the eigenvalue with least-negative real part cannot be more negative than the average eigenvalue $\lambda_i$ of $A$. Above, we have used that all eigenvalues of $A$ have negative real parts in the models considered here. The bound is saturated when the real parts of all eigenvalues of $A$ are equal, in particular when $A$ has an exceptional point of maximal order (i.e., order $m$). We can then define the benchmark rate as $\Gamma_{\text{max}}=2\abs{\Re \tr(A)}/{m}$, the factor of two follows from the homogeneous population being quadratic in $D(t)$, and compare different parameter choices by the ratio $\Gamma(t)/\Gamma_{\text{max}}$.
We mention that the spectrum of $A$ similarly constrains relaxation towards the steady state, since the environmental transient is bounded by terms linear and quadratic in the propagator norm (See App. \ref{app:environdyn}). At fixed trace and dimension, a maximal-order EP maximizes the slowest exponential decay rate of the propagator. However, its implications for a particular observable depend additionally on the contributing modes and their polynomial prefactors.  

\begin{figure}
    \centering
    \includegraphics[width=\linewidth]{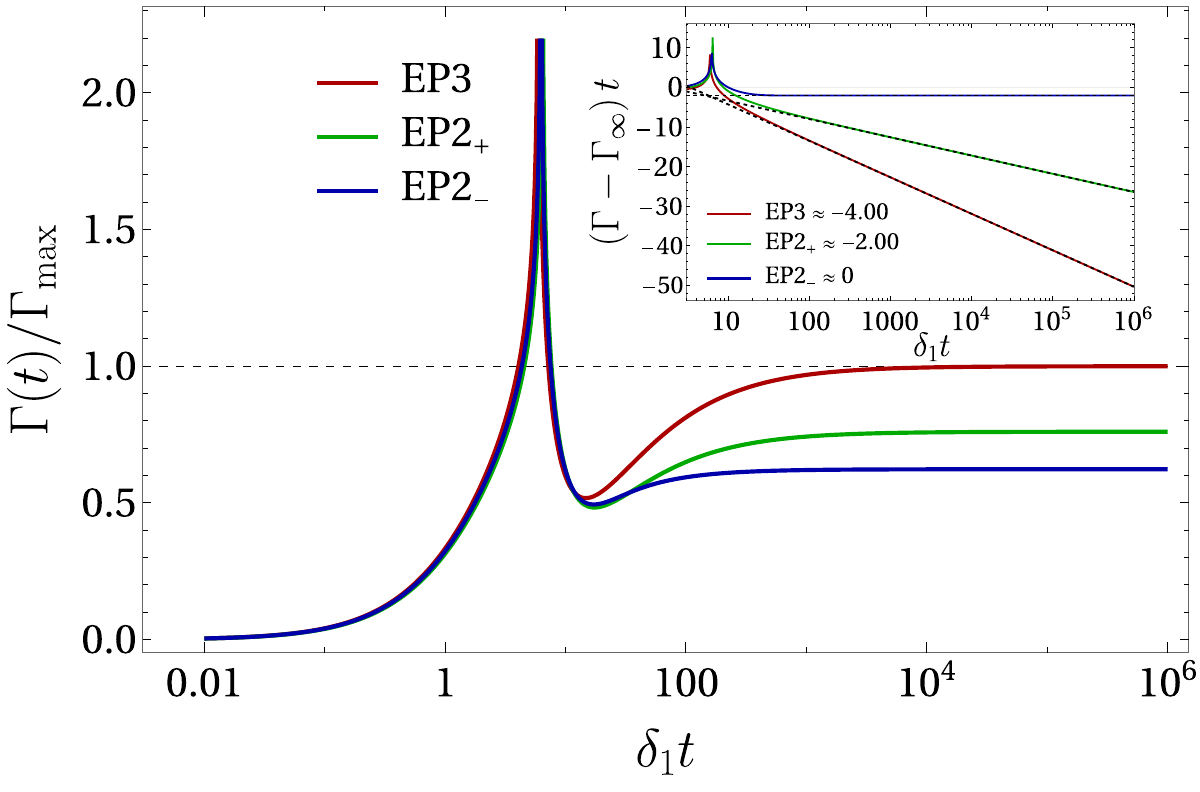}
    \caption{Rate of information loss comparing EP points of different order showing that the maximal asymptotic rate of information loss is obtained at the maximum EP order. Inset: $t[\Gamma(t)-\Gamma_\infty]$, exposing the leading polynomial prefactor of the distinguishability decay. The asymptotic slopes with respect to $\ln(\delta_1t)$ are $-4$, $-2$, and $0$ for EP3, EP2$_+$, and EP2$_-$, respectively consistent with orders of exceptional points, noting that in the case of EP2$_-$, the slowest decay is not set by the EP eigenvalue. Parameters: $\delta_1 = 1$ and $\delta_2,\kappa$ set by relation in Eq.~\eqref{eq:ep3cond} taking the minus branch. Initial dot populations: $n_d^{(1)}(0)=1$ and $n_d^{(2)}(0)=0$}
    \label{fig:dyn}
\end{figure}

Consider again the system described in Sec. \ref{sec:caseB}. With $\delta_1 \neq \delta_2$ but $\kappa_1 = \kappa_2=\kappa$, we earlier found that $A$ can exhibit a third-order EP. It follows that $\lim_{t\rightarrow\infty}\Gamma_{EP3}(t) = \Gamma_{\text{max}}=\frac{2}{3}\left(\delta_1+\delta_2\right) $. As is shown in Fig. \ref{fig:ep3}, this point lies at the intersection of two EP2 sheets. By situating ourselves at the EP3 point in parameter space and perturbing $\delta_2$ by a small amount $x$ we can then investigate the behaviour of the dynamics as it transitions from the EP3 condition to two EP2 points, which is shown in Fig. \ref{fig:dyn}. Here we perturb away from the EP3 by keeping $\delta_1$ fixed and varying $\delta_2$ by $x\delta_{1,\mathrm{EP3}}$, with $x=0.01$. This generates two new solutions for $\kappa$ which allow for EP2s labelled by $+/-$ depending on whether the new $\kappa$ is larger or smaller than the original $\kappa_{EP3}$. From Fig. \ref{fig:dyn} we see that the transient dynamics are similar, but that the long term behaviour reveals the character of the underlying EPs. For all 3 cases we observe a divergence of the rate as the two populations momentarily become equal. We see that the rate of information loss in the long time limit is maximized for the EP3 case compared to the two EP2 cases. The time required to reach this plateau is however longer, since here, the higher order EP must wait longer for the contribution from a larger polynomial prefactor to become negligible. 
Plotting $t(\Gamma(t)-\Gamma_\infty)$, where $\Gamma_\infty=\lim_{t\to\infty}\Gamma(t)$, reveals the leading polynomial prefactor, as shown in the inset of Fig. \ref{fig:dyn}. For $R(t)\sim C(\delta_1t)^p e^{-\Gamma_\infty t}$, this quantity approaches $-p\ln(\delta_1t)-\ln C$, yielding slopes of $-4, -2$, and $0$ with respect to $\ln(\delta_1t)$ for EP3, EP2$_+$, and EP2$_-$, respectively. Interestingly the EP2$_-$ branch shows a slope of $\approx 0$. Calculating the corresponding eigenvalues for this branch we find that the slowest decay mode is not associated with the defective subspace of the exceptional point, explaining the absence of a polynomial suppression in the long time limit and why its asymptotic rate of information loss is reached sooner compared to EP2$_+$.

\section{Engineered spectral densities and irreversible escape}
\label{sec:eng}

Throughout this work, we have considered sums or differences of Lorentzian spectral densities. These provide a natural starting point because they admit simple auxiliary
representations with local damping. However, non-Markovian EPs are not restricted to such spectral densities. 

Consider the model given in Ref. \cite{Alford2026}. Instead of starting with a given spectral density and determining its RC representation, let us imagine that the underlying structure consists of two tunnel-coupled RCs with energies $\omega_{1,2}$. The system couples only to RC1 with strength $\lambda$, the RCs interact with strength $g$, and only RC2 couples to a residual bath with constant spectral density
$J_{\mathrm{res}}(\omega)=\Gamma$. The joint evolution matrix is
\begin{equation}
    A=-\begin{pmatrix}
        i\epsilon_d&i\lambda&0\\
        i\lambda&i\omega_1&ig\\
        0&ig&i\omega_2+\frac{\Gamma}{2}
    \end{pmatrix}.
    \label{eq:engA}
\end{equation}

As shown in Ref. \cite{Alford2026}, the spectral density seen by the original dot is
\begin{equation}
    J(\omega)=
    \frac{\lambda^2g^2\Gamma}
    {\bigl[(\omega-\omega_1)(\omega-\omega_2)-g^2\bigr]^2
    +\frac{\Gamma^2}{4}(\omega-\omega_1)^2}.
    \label{eq:Jeng}
\end{equation}
This non-Lorentzian spectral density has a quartic denominator and can exhibit two resolved peaks associated with the hybridisation of the RCs. For distinct RC-pair eigenvalues and $t\geq0$, the retarded memory kernel is a sum of two decaying exponentials,
\begin{equation}
    K(t)=\lambda^2\sum_{\nu=\pm}c_\nu e^{-iz_\nu t},
\end{equation}
where $z_\pm$ are the eigenvalues of the lower right $2\times2$ block of $iA$ and $c_\pm$ are their generally complex weights. These complex weights reflect the coherent coupling between the RCs and distinguish this construction from sums or differences of
Lorentzian functions which lead to real weights.

The matrix in Eq.~\eqref{eq:engA} admits exceptional points, including an EP3 for positive physical parameters. Thus, a non-Lorentzian spectral density does not prevent the investigation of non-Markovian EPs through a finite-dimensional evolution matrix. More generally,
networks of coupled RCs provide a route to engineering structured spectral densities and investigating their associated exceptional points. Such geometries can lead to interesting physical properties, as we demonstrate below.

\subsection{Escape time and higher-order non-Markovian EPs}
\label{sec:esc}
Let us consider a generalisation of the above model, where a single dot (at site 1) interacts with a chain of coupled RCs (at sites $2,\dots,m$). The dot interacts with only the RC at site 2, and only the RC at site $m$ is damped. We take all site energies equal to $\epsilon_d$, all neighbouring couplings nonzero, and the terminal amplitude damping rate equal to $\Gamma/2$. Following essentially the same procedure as for two coupled RCs, it can be seen that the effective spectral density is non-Lorentzian for $m\geq 3$. Previously, we investigated information revivals associated with information returning to the dot from its environment. Now we ask a complementary question: when does an initially occupied dot lose its excitation irreversibly to the residual bath, i.e., when does the excitation leave the dot--RC system?

We assume that the dot is initially occupied, while the RCs and the residual bath are empty. Then, the probability that the excitation is still in the dot--RC system at time $t$ is given by $\mathcal P(t) = \sum_{j=1}^m |D_{j1}(t)|^2$. The probability of escape or the probability that the excitation has been irreversibly lost is simply $1-\mathcal P(t)$. With one excitation, no injection, and irreversible terminal loss, $\mathcal P(t)$ is both the total dot--RC population and the no-escape probability.
Then, the escape-time probability density is $w(t) = -\dot{\mathcal P}(t)$. We show in App. \ref{app:esc} that if the dot--RC system is at its maximal EP of order $m$, with common eigenvalue $-i\epsilon_d-\beta$, then 
\begin{align}
    w(t)=\frac{(2\beta)^{2m-1}}{(2m-2)!}\,t^{2m-2}e^{-2\beta t},
\end{align}which is an Erlang distribution, meaning a gamma distribution with integer shape $2m-1$ and rate $2\beta=\Gamma/m$. If $T$ is the random time of escape, then
\begin{equation}\label{ir:moments}
\langle T\rangle=\frac{2m-1}{2\beta},\qquad
\frac{\operatorname{Var}T}{\langle T\rangle^2}=\frac1{2m-1}.
\end{equation}Increasing the EP order reduces the fluctuations of the escape time relative to its mean. Thus, the escape time becomes more predictable; the spread of measured escape times is a smaller fraction of the average waiting time. Let us note that an Erlang distribution and $\operatorname{Var}T/\langle T\rangle^2=1/3$ have been obtained in the context of non-Hermitian Hamiltonians recently \cite{Barchielli2026}; here, we have obtained an arbitrary-order formula for non-Markovian EPs.

We next ask whether tuning to the EP can accelerate irreversible escape. To make this comparison meaningful, we fix the chain length and the sum of the squared coherent couplings, while allowing their distribution and the terminal damping to vary. Let $T_\varepsilon$ denote the time at which the remaining population falls to $\varepsilon$. We show in App.~\ref{app:esc} that, compared with any fixed non-EP configuration satisfying these constraints,
\begin{equation}
    \lim_{\varepsilon\to 0}\frac{T^{\text{Max EP}}_{\varepsilon}}{T^{\text{non-EP}}_{\varepsilon}} < 1.
\end{equation}
Thus, the maximal-order EP uniquely optimises the asymptotic escape rate under this constraint, yielding asymptotically faster escape from the dot--RC system. These results illustrate how going beyond Lorentzian spectral densities opens up additional ways to control memory and irreversible escape through the structure of the environment.

\section{Discussion}

We have introduced a Heisenberg-equation approach to non-Markovian exceptional points based on the reaction-coordinate formalism. For quadratic, number-conserving systems, it gives exact dynamics through an evolution matrix whose dimension grows linearly with the number of system and RC modes. This considerably simplifies the investigation of higher-order EPs and directly connects their spectral structure to the poles of the retarded Green function and their dynamical signatures.

Our results show that non-Markovian EPs do not generally mark the onset of information backflow. Revivals can arise from the competition between purely decaying modes, while oscillatory modes can occur on both sides of an EP curve. Different dots can also exhibit different revival counts at the same EP. Information recovery therefore depends on both the eigenvalues and their contributions to the observed subsystem. 

Nevertheless, EPs impose precise constraints on revival dynamics. We showed that a single dot coupled to any number of independent resonant Lorentzian baths exhibits exactly one revival at an EP3. In the coupled-dot model, an EP4 produces different revival counts depending on the bath widths and the observed dot. More generally, when the entire $m$-dimensional evolution matrix forms one EP, the local distinguishability for the preparations considered here has at most $m-2$ revivals. These exact bounds connect non-Markovian EPs to quantitative restrictions on information backflow and complement previous work which has connected non-Markovian revivals to the order of exceptional points \cite{Sergeev}. Higher-order non-Markovian EPs also provide a route to accelerated asymptotic relaxation; our examples demonstrate that this enhanced asymptotic decay can coexist with transient information revivals.

Finally, we used coupled RCs to investigate EPs beyond Lorentzian spectral densities and to control irreversible escape through the structure of the environment. We showed that a maximal-order EP can produce an Erlang escape-time distribution with relative fluctuations that decrease with EP order, and showed how the maximal-order EP saturates the asymptotic escape-rate bound at fixed chain length. The control of escape-time fluctuations demonstrated here
suggests exceptional-point engineering as a route to shaping tick statistics in autonomous quantum clocks~\cite{Erker2017Clocks,Schwarzhans2021Clocks,Meier2025Precision}. Extending the framework to autonomous excitation recycling and continuously driven transport could reveal how EPs influence both timing precision and correlations between successive ticks, potentially offering new routes to enhanced clock performance~\cite{Meier2026CorrelatedTicks}.

Overall, our results establish non-Markovian exceptional points as a means of controlling non-Markovian open quantum systems. The Heisenberg reaction-coordinate approach makes these effects analytically accessible and opens a route to engineering how quantum information returns to a system and how quickly it is lost.

\section*{Acknowledgements}

We thank Géraldine Haack for discussions. D.H.~acknowledges support from the Swedish Research Council Grant No. 2023-03915. G.B. acknowledges support by the Spanish State Research Agency (MCIN/AEI/10.13039/501100011033), FEDER (UE) under Grant No. PID2023-151975NB-I00. S.K.~acknowledges support from the Swiss National Science Foundation Grant No. P500PT\textunderscore222265. The authors acknowledge mathematical input from ChatGPT$^\text{\textregistered}$ 5.6 sol, specifically for the proofs presented in Apps. \ref{app:inforev} and \ref{app:esc}. The authors take full responsibility for the results and conclusions presented.

\bibliography{References}
\appendix
\onecolumngrid
\section{Model and reaction-coordinate mapping}
\label{app:rc}
We derive Eq. \eqref{eq:eomsol} presented in the main text and give a short summary of the RC method complementing \cite{Strasberg2016,nazir2018}.
\\
\\
Consider the Hamiltonian, $\hat H = \hat H_{\text{S}}+\hat H_{\text{B}} + \hat H_{\text{SB}}$, with
\begin{equation}
\hat H_\text{B} = \sum_{k=1}^{N_\text{s}}\sum_{j=1}^{N_b^k}\sum_{\alpha} \omega_{kj\alpha}\hat b^{\dagger}_{kj\alpha}\hat b_{kj\alpha}, 
    \hspace*{0.3cm}\hat H_\text{SB} = \sum_{k=1}^{N_\text{s}}\sum_{j=1}^{N_b^k}(\hat S^\dagger_{k} \hat B_{kj} + B_{kj}^\dagger\hat S_{k}),
\end{equation}
$\hat H_\text{S}$ remains unspecified and $\hat B_{kj} = \sum_\alpha t_{kj\alpha} \hat b_{kj\alpha}$. The Hamiltonian can be simplified by performing a Bogolyubov transformation. For the purposes of this work we are interested in the number conserving case, reducing the transformation to a unitary one
\begin{equation}
    \hat a_{kj\beta} = \sum_\alpha \hat{U}_{\beta \alpha}^{kj}\hat{b}_{kj\alpha}, \hspace{4pt}\text{s.t. } \hat{U}\hat{U}^\dagger =  \mathds 1.
\end{equation}
We start by fixing the mode, $\hat{a}_{kj,0}$, which becomes the corresponding RC, 
\begin{equation}
    \hat a_{kj0} = \sum_\alpha \frac{t_{kj\alpha}}{\lambda_{kj}}\hat{b}_{kj\alpha} \hspace{4pt} \text{with} \hspace{4pt} \lambda_{kj} = \sqrt{\sum_\alpha \abs{t_{kj\alpha}}^2}.
\end{equation}
As an immediate consequence we have that
\begin{equation}
    \hat{H}_\text{SB} = \sum_{k=1}^{N_\text{s}}\sum_{j=1}^{N_b^k}\lambda_{kj}\left(\hat{S}_k^\dagger \hat{a}_{kj0} + \hat{a}^\dagger_{kj0}\hat{S}_k\right). 
\end{equation}
The remaining part of the transformation, $\hat U_{n,\alpha}^{kj}$ for $n>0$ still remains undetermined, but must be orthogonal to $U_{0,\alpha}^{kj}$. 
For notational simplicity, let's denote the pseudomodes $\hat{a}_{kj0}$ by $\hat{a}_{kj}$, operators including $n$ are then understood to be taken with $n>0$.
The remaining modes can be written as $\hat{a}_{kjn} = \sum_\alpha U_{n\alpha}^{kj} \hat{b}_{kj\alpha}$. Inserting into the expression for $\hat{H}_\text{B}$ we get
\begin{equation}
    \begin{split}
        \hat{H}_{\text{B}} &=  \sum_{kj}\sum_\alpha \frac{\omega_{kj\alpha}\abs{t_{kj\alpha}}^2}{\lambda_{kj}^2}\hat{a}^{\dagger}_{kj}\hat{a}_{kj} + \sum_{kj}\sum_{\alpha,n}\left[\frac{\omega_{kj\alpha}t_{kj\alpha}}{\lambda_{kj}}U_{n\alpha}^{kj*}\hat{a}^{\dagger}_{kj}\hat{a}_{kjn} + h.c.\right] \\
        &+ \sum_{kj}\sum_{\alpha, n} \omega_{kj\alpha}\abs{U_{n,\alpha}^{kj}}^2\hat{a}^{\dagger}_{kjn}\hat{a}_{kjn} + \sum_{kj}\sum_{\alpha,n\neq n'}\omega_{kj\alpha} {U_{n,\alpha}^{kj}}U_{n',\alpha}^{kj*}\hat{a}^{\dagger}_{kjn}\hat{a}_{kjn'}.
    \end{split}
\end{equation}
The last row only contains operators for $n>0$, note that they are components of the hermitian matrix (w.r.t the $n,n'$ indices) $R_{kj}^{ nn'} = \sum_{\alpha}\omega_{kj\alpha} {U_{n,\alpha}^{kj}}U_{n',\alpha}^{kj*}$ with $n,n'\neq0$ and can therefore be diagonalized. In terms of these new diagonalized modes, $\hat{\tilde{b}}_{kjn}$ we have
\begin{equation}
    \hat{H}_\text{B} = \sum_{kj}\Omega_{kj}\hat{a}^{\dagger}_{kj}\hat{a}_{kj} + \sum_{kj}\sum_{n}(\gamma_{kjn}\hat{a}^{\dagger}_{kj}\hat{\tilde{b}}_{kjn}+\gamma_{kjn}^*\hat{\tilde{b}}^{\dagger}_{kjn}\hat{a}_{kj}) + \sum_{kjn}\Omega_{kjn}\hat{\tilde{b}}^{\dagger}_{kjn}\hat{\tilde{b}}_{kjn}.
    \label{eq:transbath}
\end{equation}
Where $\Omega_{kj} = \sum_{\alpha}\omega_{kj\alpha}\abs{t_{kj\alpha}^2}/\lambda_{kj}^2$. Note that the effect of the above transformation is to shift the environmental coupling to the system onto a single mode $\hat{a}_{kj}$ of each bath which in turn interacts with the rest of the bath (the residual modes) with a new coupling strength given by $\gamma_{kjn}$, the energies of these new modes are also renormalized by the coupling. In order to determine $\gamma_{kjn}$ we would need to consider the details of our transformation, the trick is to realize that the individual couplings are not of interest, we only need the corresponding spectral density function $\tilde J_{kj}(\omega)$, which can be determined from the original spectral density function $J_{kj}$.
\\
\\
Consider the form of our bath Hamiltonian before the transformation, for each bath indexed with $kj$ we have
\begin{equation}
    \hat{H}_\text{B}^{kj} = \bold{b}^\dagger h\bold{b} 
\end{equation}
with $\bold{b} = (\hat{b}_0,\hat{b}_1 \cdots)^T$ and $h_{\alpha,\beta} = \omega_{\alpha}\delta_{\alpha\beta}$, (here we have dropped the $kj$ indices since it's clear from the context). The unitary transform does not mix operators with different $kj$ indices, using the unitary transformation found above, we can write $\tilde{\bold{b}} = U\bold{b}$ with $\tilde{\bold{b}} = (\hat{a}_0, \hat{\tilde{b}}_1 \cdots)^T$. We then find
\begin{equation}
    \hat{H}_B^{kj} = \bold{b}^\dagger h\bold{b} = \tilde{\bold{b}}^\dagger U h  U^\dagger\tilde{\bold{b}} = \tilde{\bold{b}}^\dagger \tilde h \tilde{\bold{b}}.  
\end{equation}
We now consider the resolvent of $\tilde h$
\begin{equation}
    \tilde R(z) = (z\mathds 1 - \tilde h)^{-1} = U (z\mathds 1 - h)^{-1}U^\dagger = UR(z)U^\dagger.
    \label{eq:residual}
\end{equation}
Let's consider the component $\tilde R_{00}$,
\begin{equation}
    \tilde R_{00}(z) = \frac{1}{\lambda^2}\sum_\alpha \frac{\abs{t_\alpha}^2}{z-\omega_\alpha} = \frac{1}{2\pi\lambda^2}\int d\nu \frac{J(\nu)}{z-\nu},
\end{equation}
where we have used that $h$ is diagonal with energies $\omega_\alpha$ and the expression for $ U_{0,\alpha}$ above. Re-arranging Eq.~\eqref{eq:residual} we get 
\begin{equation}
    (z\mathds 1 - \tilde h)\tilde R(z) = \mathds 1.    
\end{equation}
Or in component form
\begin{equation}
    \sum_k (z\mathds 1 - \tilde h)_{nk}\tilde R_{km}(z) = \delta_{nm}
\end{equation}
Taking $(n,m)=(0,0)$ we get
\begin{equation}
    (z\mathds 1 - \tilde h)_{00}\tilde R_{00}(z) + \sum_{k>0} (z\mathds 1 - \tilde h)_{0k}\tilde R_{k0}(z) = 1.
\end{equation}
If we instead take $(n,m) = (>0,0)$ 
\begin{equation}
    \sum_{k} (z\mathds 1 - \tilde h)_{nk}\tilde R_{k0}(z) = 0.
\end{equation}
But from the structure of Eq. \eqref{eq:transbath}, we see that $(z\mathds 1 - \tilde h)_{nk}$ only contributes two non-zero terms
\begin{equation}
    (z\mathds 1 - \tilde h)_{nn} = z-\Omega_n, \hspace{4pt} (z\mathds 1 - \tilde h)_{n0} = -\gamma_{n}^* \implies     \tilde R_{n0}(z) = \frac{\gamma_n^*}{z-\Omega_n}\tilde R_{00}(z).
\end{equation}
Combining the expressions and simplifying we then find
\begin{equation}
\sum_{n>0}\frac{\abs{\gamma_n}^2}{z-\Omega_n} = (z-\Omega) - \frac{2\pi\lambda^2}{\int d\nu \frac{J(\nu)}{z-\nu}}.
\end{equation}
We now set $z=\omega + i\eta$ and take the limit as $\eta \rightarrow 0^+$. Taking the imaginary part gives
\begin{equation}
\Im{\lim_{\eta\rightarrow0^+}\sum_{n>0}\frac{\abs{\gamma_n}^2}{\omega+i\eta-\Omega_n}} = -\tilde J(\omega)/2.
\end{equation}
Evaluating the RHS we find
\begin{equation}
    \tilde J(\omega) = \frac{\lambda^2J(\omega)}{\left[\mathcal{P}\int\frac{d\nu}{2\pi}\frac{J(\nu)}{\omega-\nu}\right]^2+J^2(\omega)/4}.
    \label{eq:spectralupdate}
\end{equation}
Starting from a Lorentzian spectral density $J(\omega) = \frac{\kappa \delta^2}{(\omega-\omega_0)^2+\delta^2}$, Eq. \eqref{eq:spectralupdate} gives $\tilde J = 2\delta$ and $\lambda = \sqrt{\kappa\delta/2}$, i.e. starting from a Lorentzian spectral density yields a flat residual spectral density. 
\\
\\
Having derived the necessary properties of the RC method, we can now shift our attention to the Heisenberg equations. From here on we assume a quadratic, number-conserving system Hamiltonian $\hat H_\text{S} = \sum_k \omega_k \hat{S}^\dagger_k \hat S_k$. Making use of $\dot{\hat{o}}(t) = i[\hat H,\hat o(t)]$, where $\hat H = \hat H_\text{S}+\hat H_\text{SB}+ \hat H_\text{B}$, we find
\begin{align}
    &\dot{\hat{S}}_k = -i\omega_k \hat{S}_k - i \sum_{j}\lambda_{kj}\hat{a}_{kj} \label{eq:top}\\
    &\dot{\hat{a}}_{kj} = -i\Omega_{kj}\hat{a}_{kj}-i\lambda_{kj}\hat{S}_k -i\sum_{n>0}\gamma_{kjn}\hat{\Tilde{b}}_{kjn} \label{eq:middle} \\
    &\dot{\hat{\Tilde{b}}}_{kjn} = -i\Omega_{kjn}\hat{\Tilde{b}}_{kjn}-i\gamma_{kjn}^*\hat{a}_{kj} \hspace{4pt}(n>0) \label{eq:bottom}
\end{align}
Integrating Eq.~\eqref{eq:bottom} 
\begin{equation}
    \hat{\Tilde{b}}_{kjn}(t) = e^{-i\Omega_{kjn}t}\hat{\Tilde{b}}_{kjn}(0)-i\gamma_{kjn}^*\int_0^t ds e^{-i\Omega_{kjn}(t-s)}\hat{a}_{kj}(s),
\end{equation}
and inserting into Eq.~\eqref{eq:middle} gives
\begin{equation}
    \dot{\hat{a}}_{kj} = -i\Omega_{kj}\hat{a}_{kj}-i\lambda_{kj}\hat{S}_k -i\sum_{n>0}\gamma_{kjn}e^{-i\Omega_{kjn}t}\hat{\Tilde{b}}_{kjn}(0)-\sum_{n>0}|\gamma_{kjn}|^2\int_0^t ds e^{-i\Omega_{kjn}(t-s)}\hat{a}_{kj}(s).
    \label{eq:aapp1}
\end{equation}
We define the residual noise term, $\hat{\xi}_{kj}(t) = -i\sum_{n>0}\gamma_{kjn}e^{-i\Omega_{kjn}t}\hat{\Tilde{b}}_{kjn}(0)$. Focusing on the last term, this can be rewritten as
\begin{equation}
    \int d\omega \int_0^t ds \sum_{n>0}|\gamma_{kjn}|^2\delta(\omega-\Omega_{kjn})  e^{-i\omega(t-s)}\hat{a}_{kj}(s) = \frac{1}{2\pi}\int d\omega\int_0^t ds \tilde J(\omega)e^{-i\omega(t-s)}\hat{a}_{kj}(s)=\delta_{kj}\hat a_{kj}(t).
    \label{eq:aapp2}
\end{equation}
Where we have used the definition of the spectral density function and that the residual spectral density is flat when starting from a Lorentzian spectral density. Introducing $\vec{v} = (\hat{S}_1,\cdots,\hat{S}_{N_s}, \hat{a}_{11},\cdots,\hat{a}_{1N_B}, \cdots ,\hat{a}_{N_S N_B})^T$ and $\vec{\xi}(t) = (0, \cdots, 0, \hat{\xi}_{11}, \cdots, \hat{\xi}_{N_S 1}, \cdots, \hat{\xi}_{N_S N_B})^T$. Eqs.~\eqref{eq:top} and \eqref{eq:middle} together with Eqs.~\eqref{eq:aapp1} and \eqref{eq:aapp2}, give the following matrix equation
\begin{equation}
    \dot{\vec{v}} = A\vec{v}+\vec{\xi(t)}
\end{equation}
with solution
\begin{equation}
    \vec{v}(t) = D(t)\vec{v}(0) + \int_0^t ds D(t-s)\vec{\xi}(s), \hspace{8pt} D(t) = e^{At}.
\end{equation}
giving us Eq.~\eqref{eq:eomsol}, as in the main text.

\section{Properties of eigenvalues for the chain configuration}
\label{app:propeig}

\begin{figure}[t]
    \centering
    \includegraphics[width=0.2\linewidth]{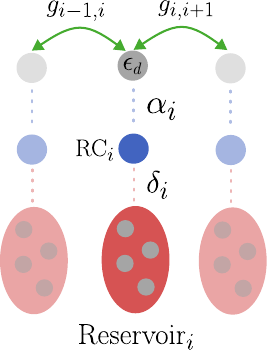}
    \caption{Schematic of chain configuration. Each dot has the common energy $\epsilon_d$ and is coupled to an individual RC with $\alpha_i,\delta_i$ as well as to its nearest neighbours through  $g_{i,i\pm 1}$}
    \label{fig:appchain}
\end{figure}
\label{app:chain}

Assuming all dots to be resonant at energy $\epsilon_d$.
The non-zero elements of the shifted evolution matrix $\tilde A=A+i\epsilon_d$ for the chain configuration (depicted in Fig. \ref{fig:appchain}) are given by 
\begin{equation}
    \begin{aligned}
        &\tilde A_{ii} =-\delta_i, \quad i=N_d+1,\dots,2N_d\\
        &\tilde A_{i+1,i} = \tilde A_{i,i+1} = -ig, \quad i=1,\dots,N_d-1\\
        &\tilde A_{N_d+i,i} = \tilde A_{i,N_d+i} = -i\alpha_i,\quad i =1,\dots,N_d
    \end{aligned}
\end{equation}
where $N_d$ is the number of quantum dots in the chain. In this form $\tilde A$ has a convenient block matrix structure in terms of the real matrices $G,\Lambda$ and $\Delta$
\begin{equation}
    \tilde A =-\begin{pmatrix}
        iG & i\Lambda \\
        i\Lambda & \Delta
    \end{pmatrix}
\end{equation}
where $\Lambda = \text{diag}\{\alpha_1,...,\alpha_{N_d}\}$, $\Delta = \text{diag}\{\delta_{N_d+1},...,\delta_{2N_d}\}$ and $G_{i,j} = g\delta_{i,j\pm 1}$ with $i,j\in \{1,...,N_d\}.$ Introducing the diagonal sign matrix $Q_{i,j} = (-1)^{i}\delta_{i,j}$ we note that $QGQ = -G$ since $(QGQ)_{ij} = g(-1)^{i+j}\delta_{i,j\pm1} = -g \delta_{i,j\pm1}$ and similarly $Q\Delta Q = \Delta$ and $Q \Lambda Q = \Lambda$. By means of $Q$ the vector space within each block is split into two subspaces with positive and negative parity. We can then define the projectors $P_{\pm} = \frac{1}{2}(\mathds 1 \pm Q)$ onto these subspaces (Note $P_{\pm}^2 = P_{\pm}$ and $P_+P_- = 0, P_++P_- = \mathds 1$). We then have $P_\pm G = GP_\mp$ while $P_\pm \Delta = \Delta P_\pm$ and $P_\pm \Lambda = \Lambda P_\pm$. The similarity transform
\begin{equation}
    S = \begin{pmatrix}
        P_+ + iP_- & 0 \\
        0 & iP_+ + P_-
    \end{pmatrix},
    \hspace{4pt}
    S^{-1} = \begin{pmatrix}
                P_+ -iP_- & 0 \\
        0 & -iP_+ + P_-
    \end{pmatrix}
\end{equation}
then gives
\begin{equation}
    S\tilde A S^{-1} =  \begin{pmatrix}
        GQ & -\Lambda Q \\
        \Lambda Q & -\Delta
    \end{pmatrix}.
\end{equation}
Since $\tilde A$ is similar to a real matrix its characteristic polynomial must have real coefficients and its roots must come in complex conjugate pairs. Note that we above do not need to make any assumption about the values of the couplings between the dots $g_{ij}$ apart from the fact that $G$ must be a matrix that anti-commutes with $Q$, which is for instance not the case when generalizing the nearest neighbour coupling to an all-to-all coupling. In general the same construction can be applied to show that $\tilde A$ is similar to a real matrix when the spectrum of $G$ is symmetric (taking multiplicity of eigenvalues into account) and both $\Delta, \Lambda$ commute with $Q$. We stress that this is only a sufficient and not a necessary condition for $\tilde A$ to have a characteristic polynomial with real coefficients. Interestingly, it can be shown that a fully connected chain of dots with all $g_{ij}>0$ can never generate a matrix $\tilde A$ that is similar to a real matrix when $N_d \geq 3$.

\section{BLP measure}
\label{app:blp}
Here we derive Eq. \eqref{eq:nblp}. Since the EP and overdamped situations involve monotonic decay, these cases lead to $\mathcal N_{\text{BLP}}=0$. We focus only on the underdamped regime. Note that for the single-level quantum dot model, the time derivative of the distinguishability is given by
\begin{align}\label{eq:dera}
    \frac{d\mathcal D(t,n_d^{(1)},n_d^{(2)})}{dt} = 2\frac{e^{-\delta t}\alpha^2}{\eta'^2}\bigg(\delta \cos(\eta' t)-\eta'\sin(\eta' t)-\delta \bigg)
\end{align}The derivative being positive is equivalent to 
\begin{equation}
    \begin{aligned}
        \delta \cos(\eta't) - \eta'\sin(\eta't) - \delta > 0\\
        \implies\delta(1-\cos(\eta't))+\eta'\sin(\eta't) <0.
    \end{aligned}
\end{equation}Using half-angle identities and defining $R = \sqrt{\delta^2+\eta'^2}$ and $\phi=\tan^{-1}(\eta'/\delta)$, the above inequality is equivalent to
\begin{align}
    R\sin\left(\frac{\eta't}{2} \right)\sin\left(\frac{\eta't}{2}+\phi \right) <0.
\end{align}This is satisfied for all values of time ($k\in \mathds N$) such that \begin{align}
    \frac{2}{\eta'}(k\pi-\phi) <t<\frac{2k\pi}{\eta'}.
\end{align}
Therefore, the derivative {in Eq.}~\eqref{eq:dera} must be integrated over all such intervals to obtain $\mathcal N_{\text{BLP}}$. It is then simple to obtain that (also using $\eta'=\sqrt{4\alpha^2-\delta^2}$)
\begin{align}
    \mathcal N_{\text{BLP}} = \sum_{k=1}^{\infty} e^{-\frac{2k\pi\delta}{\eta'}} = \frac{1}{e^{\frac{2\pi\delta}{\eta'}}-1}.
\end{align}

\section{Case B}
\label{app:ep3}

\subsection{Eigenvalues and EPs}
\begin{align}
p &= \frac{\delta_1\delta_2+\alpha_1^2+\alpha_2^2}{3}
     -\frac{(\delta_1+\delta_2)^2}{9},\\
q &= -\frac{(\delta_1+\delta_2)^3}{27}
     +\frac{(\delta_1+\delta_2)
     (\delta_1\delta_2+\alpha_1^2+\alpha_2^2)}{6}
     -\frac{\alpha_1^2\delta_2+\alpha_2^2\delta_1}{2}.
\end{align}Below, we give the three algebraic branches of the condition
$\Delta_{\mathrm B}=0$ discussed in the main text.
Using $\alpha_i^2=\kappa\delta_i/2$ and defining
$s=\delta_1+\delta_2$ and $t=\delta_1\delta_2$, this condition
becomes
\begin{equation}
-\frac{s^3}{2}\kappa^3
+\left(\frac{s^4}{4}+6s^2t-27t^2\right)\kappa^2
-3st(s^2-4t)\kappa+t^2(s^2-4t)=0.
\end{equation}
Dividing by $-s^3/2$, we obtain
$\kappa^3+b\kappa^2+c\kappa+d=0$, where
\begin{equation}
b=-\frac{s^4+24s^2t-108t^2}{2s^3},
\qquad
c=\frac{6t(s^2-4t)}{s^2},
\qquad
d=-\frac{2t^2(s^2-4t)}{s^3}.
\end{equation}
Introducing
\begin{equation}
\begin{aligned}
P&=\frac{c}{3}-\frac{b^2}{9},
& Q&=-\frac{b^3}{27}+\frac{bc}{6}-\frac{d}{2},\\
U&=\sqrt[3]{Q+\sqrt{Q^2+P^3}},
& V&=\sqrt[3]{Q-\sqrt{Q^2+P^3}},
\end{aligned}
\end{equation}
with cube-root branches chosen such that $UV=-P$,
the three solutions are
\begin{equation}
\begin{aligned}
\kappa_{\mathrm{EP}}^{(1)}&=-\frac{b}{3}+U+V,\\
\kappa_{\mathrm{EP}}^{(2)}&=-\frac{b}{3}+\omega U+\omega^2V,\\
\kappa_{\mathrm{EP}}^{(3)}&=-\frac{b}{3}+\omega^2U+\omega V,
\end{aligned}
\qquad \omega=e^{2\pi i/3}.
\end{equation}
Only real, positive solutions are physically admissible.
For $\delta_1\neq\delta_2$, these yield EP2s unless
$p=q=0$, where an EP3 occurs.
For $\delta_1=\delta_2=\delta$, the cubic reduces to
$\kappa^2(\kappa-\delta/4)=0$.
The positive solution $\kappa=\delta/4$ is an EP2,
whereas $\kappa=0$ is an ordinary degeneracy.

\subsection{Sufficient condition for non-Markovian revivals}
For case B, a complex-conjugate pair necessarily produces revivals. To see this, define the homogeneous dot amplitude $u(t)=[e^{\tilde A t}]_{11}$, which quantifies the decay of the initial dot population (see Eq. \eqref{eq:eomsol1}). For initially
occupied and empty dot states coupled to the same initially uncorrelated baths, the trace distance is $|u(t)|^2$. First consider unequal widths, taking $\delta_1<\delta_2$ without loss of generality. Writing the eigenvalues as $r$ and $a\pm ib$, one has
\begin{equation}
    u(t)=c_r e^{rt}
    +e^{at}\bigl[C\cos(bt)+S\sin(bt)\bigr],
    \qquad
    c_r=\frac{(r+\delta_1)(r+\delta_2)}
    {(r-a)^2+b^2}<0.
\end{equation}
Here, the negativity follows from
$-\delta_2<r<-\delta_1$ for $\delta_1<\delta_2$.
If the complex pair decays more slowly, it forces oscillatory sign changes; if the real mode decays more slowly, its negative coefficient forces a sign change from the initial positive amplitude. Equal decay rates likewise force a sign change. In every case, the resulting zero crossing is followed by a revival of $\mathcal D(t)$. Thus, $\Delta_{\text{B}}<0$ guarantees information
backflow.
For equal widths $\delta_1=\delta_2=\delta$, the real mode is dark with respect to the dot: $r=-\delta$ and $c_r=0$. The dot amplitude then reduces to the single-Lorentzian result with an enhanced effective coupling, as discussed in Sec.~\ref{sec:caseB}. Thus, a complex-conjugate pair also implies revivals in this case.
\\
\\
However, $\Delta_{\text{B}}>0$ is not a guarantee for the absence of information backflow, this can be surmised by exploiting the fact that the Laplace transform of the matrix exponential is given by its resolvent.
In the following argument, we consider the resonant case with $\kappa_1=\kappa_2=\kappa>0$, positive linewidths, and $\Delta_{\text{B}}>0$, so that the shifted eigenvalues are real and distinct. We suppress the tildes below: $A$, $\lambda_i$, and $D(t)$ denote the shifted matrix $\tilde A$, its eigenvalues, and its propagator, respectively.
We have
\begin{equation}
    \mathcal{L}(e^{At})(s) = (s\mathds 1-A)^{-1}.
\end{equation}
Using the inverse formula in terms of the adjugate matrix we have
\begin{equation}
     (s\mathds 1-A)^{-1} = \frac{\text{Adj}(s\mathds1-A)}{\mathbb P(s)}
\end{equation}
where $\mathbb P(s)$ is the determinant of $s\mathds1-A$, i.e. the characteristic polynomial of $A$.
Performing a partial fraction expansion we get
\begin{equation}
    (s\mathds 1-A)^{-1} = \sum_i \frac{C_i}{s-\lambda_i}
\end{equation}
where $C_i = \text{Adj}(\lambda_i\mathds 1 - A)/\mathbb P'(\lambda_i)$ and $\lambda_i$ are the eigenvalues of $A$. Taking the inverse Laplace transform now yields 
\begin{equation}
    \mathcal{L}^{-1}((s\mathds 1-A)^{-1})(t) = \sum_i \frac{\text{Adj}(\lambda_i\mathds 1 - A)}{\mathbb P'(\lambda_i)}e^{\lambda_i t} = D(t).
\end{equation}
Investigating the decay properties is now easily performed in Laplace space, using $D(0) = \mathds 1$ we get
\begin{equation}
    \mathcal{L}(D'(t))(s) = s\frac{\text{Adj}(s\mathds1-A)}{\mathbb P(s)} - \mathds 1.
\end{equation}
We are interested in the properties of $D'(t)_{11}$. Away from exceptional points of $A$ we can label the eigenvalues in order of their real part s.t. $0 > \Re{\lambda_1} > \Re{\lambda_2} > \Re{\lambda_3}$. Evaluating the adjugate matrix at the corresponding index and simplifying yields
\begin{equation}
    \mathcal{L}(D'_{11}(t))(s) = -\frac{s\frac{\kappa(\delta_1+\delta_2)}{2} + \kappa \delta_1\delta_2}{P(s)} = -\frac{\kappa(\delta_1+\delta_2)}{2(s-\lambda_2)(s-\lambda_3)} - \frac{\kappa \delta_1\delta_2 + \frac{\kappa(\delta_1+\delta_2)}{2}\lambda_1}{P(s)},
\end{equation}
where we have added and subtracted $\lambda_1$. Applying the inverse Laplace transform we find
\begin{equation}
    D'_{11}(t) = \frac{\kappa(\delta_1+\delta_2)}{2(\lambda_2-\lambda_3)}\left(e^{\lambda_3 t}- e^{\lambda_2 t}\right) - \left(\kappa\delta_1\delta_2+\frac{\kappa(\delta_1+\delta_2)}{2}\lambda_1\right)\int_{0}^td\tau\frac{\left( e^{\lambda_2 \tau}-e^{\lambda_3 \tau}\right)}{\left(\lambda_2-\lambda_3\right)}e^{\lambda_1(t-\tau)}.
\end{equation}
Above we have used that the product of functions in $s-$space yields a convolution in $t-$space. For $\Delta_\text{B} > 0$ all $\lambda_i$ are real and distinct. Since $\lambda_3 < \lambda_2$ the first term is negative for all $t>0$. The second term consists of a convolution between  positive functions for all $t$ and a prefactor, the overall derivative is then negative as long as 
\begin{equation}
    \abs{\lambda_1} \leq \frac{2\delta_1\delta_2}{(\delta_1+\delta_2)}.
\end{equation}
At long times the second term will dominate the derivative, since it contains the slowest decaying eigenvalue. Since $D_{11}(t)\rightarrow 0$ and $D_{11}(0)=1$, the steady state $0$ must be approached from below when the above criterion is violated. Since $D_{11}(t)$ must then have crossed zero at some time $t$, this forces a revival of the trace distance and a non-zero BLP. This can easily be engineered while keeping $\Delta_B > 0$. Thus, even with purely real eigenvalues this model can show non-Markovian revivals.

\section{Environment dynamics}
\label{app:environdyn}
Given $D(t) = e^{At}$ the environmental component of the dynamics can be simplified by making use of the resolvent of $A$. Throughout the calculations below we make use of the fact that the real parts of the eigenvalues of $A$ are all strictly negative. We have
\begin{equation}
    K(t,\omega) = \int_0^t ds D(s)e^{i\omega s} = \int_0^t ds e^{(A+i\omega \mathds 1)s}
    = R(\omega)\left(\mathds 1 -e^{(A+i\omega \mathds 1)t}\right),
\end{equation}
where $R(\omega)$ is the resolvent, $R(\omega) = \left(-i\omega \mathds 1 - A\right)^{-1}$. The environmental contribution to the population of mode $n$, assuming the residual spectral densities are flat and the initial system–environment state is uncorrelated, is given by
\begin{equation}
    E_n(t) = \sum_{k\neq n}\frac{\delta_k}{\pi}\int_{-\infty}^{\infty} \int_{0}^t\int_0^t d\omega ds ds' D_{n,k}^*(t-s)D_{n,k}(t-s')e^{-i(s'-s)\omega}f_k(\omega),
\end{equation}
making the change of variables $t-s=u, t-s'=v$ we can recast this expression using the formula for $K(t,\omega)$ as 
\begin{equation}
    E_n(t) = \sum_{k\neq n} \frac{\delta_k}{\pi}\int_{-\infty}^{\infty}d\omega\abs{K_{n,k}(t,\omega)}^2f_k(\omega).
    \label{eq:environkernel}
\end{equation}
The steady-state for population $n$ can then be directly obtained as
\begin{equation}
    E_n(\infty) = n_{n,ss} =  \sum_{k\neq n} \frac{\delta_k}{\pi}\int d\omega \abs{R_{n,k}(\omega)}^2 f_k(\omega).
\end{equation}
Define $\Gamma$ as the real part of the diagonal of $-A$, such that it fulfils the relation $A+A^\dagger=-2\Gamma$. The matrix $X=\mathds 1$ then
solves the Lyapunov equation
$AX+XA^\dagger=-2\Gamma$.
As all eigenvalues of $A$ have strictly negative real parts,
this solution is unique and satisfies
\begin{equation}
X=\int_0^\infty dt\,e^{At}(2\Gamma)e^{A^\dagger t}
 =\frac{1}{\pi}\int_{-\infty}^{\infty}d\omega\,
 R(\omega)\Gamma R^\dagger(\omega)
 =\mathds 1.
\end{equation}
Using $0\leq f_k(\omega)\leq1$ and $\delta_k\geq0$, we obtain
\begin{equation}
0\leq n_{n,\mathrm{ss}}
\leq \frac{1}{\pi}\int_{-\infty}^{\infty}d\omega\,
\left[R(\omega)\Gamma R^\dagger(\omega)\right]_{nn}
=1.
\end{equation}
\\
\\
Using Eq.~\eqref{eq:environkernel}
the environmental contribution to the evolution of population $n$ can be written as
\begin{equation}
E_n(t) = \sum_{k\neq n} \frac{\delta_k}{\pi}\int_{-\infty}^{\infty}d\omega\abs{K_{n,k}(t,\omega)}^2f_k(\omega) = \frac{1}{\pi}\int d\omega \left(K(t,\omega)\Gamma(\omega)K^\dagger(t,\omega)\right)_{nn},
\end{equation}
where the definition of $\Gamma$ has been extended to include the corresponding Fermi function. Inserting the definition of $K(t,\omega)$ we have
\begin{align}
    E_n(t) &= n_{n,ss} + \left\{D(t)\left[\frac{1}{\pi}\int d\omega R(\omega)\Gamma(\omega)R^\dagger(\omega)\right]D^\dagger(t)\right\}_{n,n} \\
    -&\left\{D(t)\left[\frac{1}{\pi}\int d\omega R(\omega)\Gamma(\omega)R^\dagger(\omega)e^{i\omega t}\right]\right\}_{n,n} -\left\{\left[\frac{1}{\pi}\int d\omega R(\omega)\Gamma(\omega)R^\dagger(\omega)e^{-i\omega t}\right]D^\dagger(t)\right\}_{n,n}. 
\end{align}
Further simplifying by defining $Q(t) = \frac{1}{\pi}\int d\omega R(\omega)\Gamma(\omega)R^\dagger(\omega)e^{i\omega t}$, gives
\begin{align}
    E_n(t) &= n_{n,ss} + \left\{D(t)Q(0)D^\dagger(t)\right\}_{n,n} 
    -\left\{D(t)Q(t)\right\}_{n,n} -\left\{Q^\dagger(t)D^\dagger(t)\right\}_{n,n}.
\end{align}
$\abs{E_n(t)-n_{n,ss}}$ is then bounded by
\begin{align}
    &\abs{\abs{D(t)Q(0)D^\dagger(t) 
    -D(t)Q(t) -Q^\dagger(t)D^\dagger(t)}} \\
    &\leq \abs{\abs{D(t)Q(0)D^{\dagger}(t)}} + \abs{\abs{D(t)Q(t)}}+\abs{\abs{Q^\dagger(t)D^{\dagger}(t)}} \\
    &\leq \abs{\abs{Q(0)}}\left(\abs{\abs{D(t)}}^2+2\abs{\abs{D(t)}}\right).
\end{align}
Since $\abs{\abs{D(t)}}$ is bounded by the slowest decaying mode, we see that the transient dynamics induced by the environment is similarly bounded by the slowest decaying mode. The environmental contribution introduces no asymptotic exponential relaxation timescale longer than that set by $A$. Consequently, the environmental contribution introduces no slower asymptotic exponential envelope than that set by $A$. For a fixed trace, the maximal-order EP then maximizes decay rate of the slowest decaying mode.

\section{Case C}

\subsection{No EPs of order beyond three for a dot coupled with resonant Lorentzian baths}\label{app:ep4pr}

We consider a single quantum dot coupled with $N_b$ independent Lorentzian baths. We show that the corresponding evolution matrix $A$ cannot have any EP of order greater than three.

First suppose that the widths $\delta_j$ are distinct. Define the characteristic polynomial $P$ and the polynomial $Q$ containing the RC diagonal entries by $ P(z)=\det(z\mathds{1}-\tilde{A})$ and $Q(z)=\prod_{j=1}^{N_b}(z+\delta_j)$. Eliminating the RC components in $z\mathds{1}-\tilde{A}$ gives
\begin{equation}
 \frac{P(z)}{Q(z)}=z+\sum_{j=1}^{N_b}\frac{\alpha_j^2}{z+\delta_j}.
 \label{eq:ratio}
\end{equation}A zero of $P/Q$ is an eigenvalue of $\tilde A$, with the same multiplicity as the corresponding zero of $P$. Between any two adjacent poles on the real axis, $P/Q$ starts at $+\infty$ and ends at $-\infty$; the term associated with the left pole diverges positively just to its right, and the term associated with the right pole diverges negatively just to its left. Continuity between the poles therefore guarantees at least one real zero in each of the $N_b-1$ intervals. Thus $P$ has at least $N_b-1$ distinct real roots. Since $P$ has real coefficients, every nonreal root has its complex conjugate as a root of the same multiplicity. A repeated nonreal root would therefore require at least four nonreal roots counted with multiplicity. Together with the $N_b-1$ distinct real roots already found, this would require at least $N_b+3$ roots. This is impossible because $P$ has degree $N_b+1$. Hence every repeated eigenvalue is real.

At a real point away from the poles, differentiating Eq.~\eqref{eq:ratio} three times gives
\begin{equation}
 \frac{d^3}{dz^3}\frac{P(z)}{Q(z)}
 =-6\sum_{j=1}^{N_b}\frac{\alpha_j^2}{(z+\delta_j)^4}<0.
 \label{eq:third}
\end{equation}
A zero of multiplicity four or greater would make the function and its first three derivatives vanish.  Eq.~\eqref{eq:third} obviously does not satisfy this. Every repeated eigenvalue therefore has multiplicity at most three. Therefore an EP of order greater than 3 is impossible for this model.

This conclusion can be easily extended to situations with some widths equal. If several RCs have the same width, only the combination proportional to their couplings interacts with the dot. They can therefore be replaced by one effective RC whose squared coupling is the sum of their squared couplings. The remaining combinations are independent, decoupled modes and cannot increase the EP order. The argument for distinct widths thus also covers equal widths.

\subsection{Band-gap structure}
\label{app:bandgap}

We derive the coefficients entering the eigenvalues
$\lambda=\alpha\pm\sqrt{\beta\pm\sqrt{\gamma}}$
of Eq.~\eqref{eq:bandgap}. 
The four eigenvalues of $A$ are
\begin{equation}
\lambda_{\sigma,\tau}
=\alpha\pm\sqrt{\beta\pm\sqrt{\gamma}},
\qquad 
\end{equation}
where
\begin{align}
\alpha={}&-i\epsilon_d-\frac{2\delta_1+3\delta_2}{8},
\\
\beta={}&
\frac{4\delta_1^2-12\delta_1\delta_2
+13\delta_2^2-16\delta_2\kappa_2}{64},
\\
\gamma={}&\frac{\delta_2^2}{256}
\Bigl[
16\kappa_2^2+(48\delta_1-40\delta_2)\kappa_2
+4\delta_1^2-12\delta_1\delta_2+9\delta_2^2
\Bigr].
\end{align}
The physical parameter region additionally requires
$\kappa_1>0$, i.e.,
$4\delta_1^2-\delta_2^2+4\delta_2\kappa_2>0$.

The discriminant of this biquadratic polynomial is
\begin{equation}
\Delta_{\mathrm{BG}}
=256\gamma^2(\beta^2-\gamma).
\end{equation}
Thus, repeated roots occur both at $\gamma=0$,
which gives two double roots when $\beta\neq0$,
and at $\beta^2=\gamma$, which gives a double root
at $y=0$ when $\beta\neq0$.
A quadruple root requires $\beta=\gamma=0$. EPs can be determined by additionally checking the merging eigenvectors.

Solving $\gamma=0$ gives
\begin{equation}
\frac{\kappa_2}{\delta_2}
=-\frac32\frac{\delta_1}{\delta_2}
+\frac54
\pm\sqrt{
2\frac{\delta_1^2}{\delta_2^2}
-3\frac{\delta_1}{\delta_2}+1},
\end{equation}
recovering Eq.~\eqref{eq:bgEP2}.
Within the positive-parameter region, imposing also
$\beta=0$ gives
$\delta_1=\kappa_2=\delta_2/2$.

\section{Information revivals at EPs}
\label{app:inforev}
\subsection{EP3}
\label{app:71}
Consider one dot coupled to $N_b$ independent Lorentzian baths
centred at $\epsilon_d$. %, with $\delta_j>0$ and $\alpha_j^2=\kappa_j\delta_j/2>0$. 
We prove that any EP3 gives exactly
one revival of the
distinguishability. In the dot--RC ordering, $\tilde A=A+i\epsilon_d\mathds{1}$ has the dot diagonal entry zero,
RC diagonal entries $-\delta_j$, and dot--RC entries $-i\alpha_j$ in both
positions. We set $u(t)=[e^{\tilde{A} t}]_{11}=e^{i\epsilon_dt}D_{11}(t)$, so that $\mathcal D(t,\rho_1,\rho_2)=u(t)^2$, where we have noted that $u(t)$ is real. We further have
\begin{equation}
	u(0)=1,\qquad \dot u(0)=0,\qquad
	\ddot u(0)=-\sum_j\alpha_j^2<0.
	\label{ir:initial-three}
\end{equation}
As before, it is simple to see that RCs with the same width can be combined into one effective RC whose squared coupling to the dot is the sum of their squared couplings. This leaves the dot amplitude unchanged, so we may assume distinct RC widths in the proof. 

\subsubsection{Eigenvalues and amplitude coefficients}
 We use a similar approach as in the previous appendix. The characteristic polynomial $P(z)$ of $\tilde A$ satisfies
\begin{equation}
	\frac{P(z)}{Q(z)}
	=z+\sum_j\frac{\alpha_j^2}{z+\delta_j},
	\qquad
	P(z)=\det(z\mathds{1}-\tilde{A}),\quad Q(z)=\prod_j(z+\delta_j).
	\label{ir:characteristic-three}
\end{equation}As before, $P$ has degree $N_b+1$, and at least $N_b-1$ distinct real roots. Since $P$ has real coefficients, a non-real triple root would require a conjugate triple root. Together with the $N_b-1$ real roots guaranteed between adjacent poles, this would exceed its degree $N_b+1$. The EP3 eigenvalue is therefore real and must lie in one of these intervals; otherwise, its threefold multiplicity would again exceed the available number of roots. Counting multiplicities then leaves exactly one simple root in every other interval and no roots elsewhere. Thus all eigenvalues are real. They are also strictly negative, since for real $z\geq0$,
\[
\frac{P(z)}{Q(z)}
=z+\sum_j\frac{\alpha_j^2}{z+\delta_j}>0,
\]
so no eigenvalue can lie in that range.
At each remaining simple eigenvalue $\tilde\lambda_\ell$, $P/Q$ crosses from positive to negative, so $\left.(P/Q)'\right|_{z=\tilde\lambda_l}<0$. We write the EP3 eigenvalue as
$-\beta$, with $\beta>0$.

We now determine the contribution of each simple eigenvalue to the dot amplitude $u(t)$. At the EP3, $u(t)=e^{\tilde\lambda_{\text{EP}}t}(b_0+b_1t+\frac{1}{2}b_2t^2)
+\sum_\ell c_\ell e^{\tilde\lambda_\ell t}$,
where the quadratic polynomial arises from the three coalescing modes, and $c_\ell$ is the coefficient associated with the remaining simple eigenvalue $\tilde\lambda_\ell$. We show below that every $c_\ell$ is negative. Let $\boldsymbol\rho_\ell$ be the right eigenvector of $\tilde A$ associated with the simple eigenvalue $\tilde\lambda_\ell$. Choosing its dot component to be 1, the RC equations give $(\rho_\ell)_{{\text{RC}},j}
={-i\alpha_j}/({\tilde\lambda_\ell+\delta_j})$. Since $\tilde A^T=\tilde A$, its transpose $\boldsymbol\rho_\ell^T$ is a left eigenvector. Consequently, the contribution of this eigenvalue to the matrix exponential is $e^{\tilde\lambda_\ell t}
{\boldsymbol\rho_\ell\boldsymbol\rho_\ell^T}/
{\boldsymbol\rho_\ell^T\boldsymbol\rho_\ell}$. Taking the dot-dot entry gives its contribution $c_\ell e^{\tilde\lambda_\ell t}$ to $u(t)$, with
\begin{equation}
c_\ell
=\frac{1}{\boldsymbol\rho_\ell^T\boldsymbol\rho_\ell}
=\left[
1-\sum_j\frac{\alpha_j^2}
{(\tilde\lambda_\ell+\delta_j)^2}
\right]^{-1}
=\frac{1}{(P/Q)'(\tilde\lambda_\ell)}<0.
\label{ir:coeff-three}
\end{equation}

The initial conditions fix $b_0$ and $b_1$. We next determine the sign of the quadratic coefficient $b_2$, which controls the long-time behaviour of the EP3 contribution. The triple-root conditions from Eq.~\eqref{ir:characteristic-three} are
\begin{equation}
	\sum_j\frac{\alpha_j^2}{\delta_j-\beta}=\beta,\qquad
	\sum_j\frac{\alpha_j^2}{(\delta_j-\beta)^2}=1,\qquad
	\sum_j\frac{\alpha_j^2}{(\delta_j-\beta)^3}=0.
	\label{ir:conditions-three}
\end{equation}
Introduce three auxiliary vectors, an eigenvector and its two
generalised eigenvectors, by their dot and RC components,
\begin{equation}
	\begin{array}{c|cc}
		&{\rm dot}&{\rm RC},\ j\\ \hline
		\boldsymbol v_0&1&-i\alpha_j/(\delta_j-\beta)\\
		\boldsymbol v_1&0&i\alpha_j/(\delta_j-\beta)^2\\
		\boldsymbol v_2&0&-i\alpha_j/(\delta_j-\beta)^3
	\end{array}
\end{equation}which satisfy $(\tilde{A}+\beta\mathds{1})\boldsymbol v_0=0$, $(\tilde{A}+\beta\mathds{1})\boldsymbol v_1=\boldsymbol v_0$ and $(\tilde{A}+\beta\mathds{1})\boldsymbol v_2=\boldsymbol v_1$. To calculate $u(t)$, we expand the vector $\boldsymbol e_1=(1,0,\ldots,0)^T$ as $\boldsymbol e_1=b_0\boldsymbol v_0+b_1\boldsymbol v_1+b_2\boldsymbol v_2
+\sum_\ell c_\ell\boldsymbol\rho_\ell$, apply $e^{\tilde A t}$ and take the dot component, 
\begin{equation}
	u(t)=e^{-\beta t}(b_0+b_1t+\tfrac12b_2t^2)
	+\sum_\ell c_\ell e^{\tilde\lambda_\ell t}.
	\label{ir:EP3amplitude}
\end{equation}
In particular, the $t^2/2$ term comes from
$e^{\tilde{A} t}\boldsymbol v_2=e^{-\beta t}
(\boldsymbol v_2+t\boldsymbol v_1+t^2\boldsymbol v_0/2)$. To determine the sign of $b_2$, we multiply the initial-vector expansion by $\boldsymbol v_0^T$.
Since $\boldsymbol v_0^T\boldsymbol v_0=\boldsymbol v_0^T\boldsymbol v_1=0$ by
Eq.~\eqref{ir:conditions-three}, and the eigenvalues are distinct, we have
$\boldsymbol v_0^T\boldsymbol\rho_\ell=0$.
We therefore have,
\begin{equation}
	1=b_2\boldsymbol v_0^T\boldsymbol v_2
	=-b_2\sum_j\frac{\alpha_j^2}{(\delta_j-\beta)^4},
	\qquad b_2<0.
	\label{ir:quadratic-three}
\end{equation}
Note that all coefficients in Eq.~\eqref{ir:EP3amplitude} are real and the values
of $b_0,b_1$ are not needed below. 

\subsubsection{Counting zeros and revivals}

Consider the rescaled amplitude $e^{\beta t}u(t)$. Eqs.~\eqref{ir:coeff-three} and \eqref{ir:quadratic-three} imply
\begin{align}
	\frac{d^2}{dt^2}[e^{\beta t}u(t)]
	&=b_2+\sum_\ell c_\ell(\tilde\lambda_\ell+\beta)^2
	e^{(\tilde\lambda_\ell+\beta)t}<0,\\
	\frac{d^2}{dt^2}[e^{\beta t}\dot u(t)]
	&=-\beta b_2+\sum_\ell c_\ell\tilde\lambda_\ell
	(\tilde\lambda_\ell+\beta)^2
	e^{(\tilde\lambda_\ell+\beta)t}>0.
\end{align}
Thus the rescaled amplitude is strictly concave. It starts at one
with slope $\beta>0$ and tends to $-\infty$; either its negative
quadratic term dominates, or a slower exponential with negative
coefficient does. Its strictly decreasing slope therefore gives
one maximum followed by exactly one positive-time zero $t_0$,
with $\dot u(t_0)<0$.

To exclude additional revivals, we also need the zeros of $\dot u$.
The rescaled derivative is strictly convex, starts at zero with
negative slope $-\sum_j\alpha_j^2$, and tends to $+\infty$;
its quadratic coefficient is positive, as is the coefficient
$c_\ell\tilde\lambda_\ell$ of any growing exponential.
It therefore has one minimum followed by exactly one positive-time
zero $t_1$. Since $\dot u(t_0)<0$, $t_1>t_0$.

Consequently, $u$ decreases through zero, reaches one negative
minimum, and then increases towards zero. The distinguishability
obeys $\dot{\mathcal D}=2u\dot u>0$ only for $t_0<t<t_1$.
It has exactly one positive-time zero and exactly one revival,
whether or not the EP3 is the slowest decay eigenvalue.

\subsection{EP4 in coupled dots}
\label{app:72}
\subsubsection{EP4 conditions and dot amplitudes}
Consider the shifted evolution matrix of the coupled dot setup in Eq.~\eqref{eq:coupled}. The characteristic polynomial is
\begin{equation}\label{ir:quartic}
P(s)=[s(s+\delta_1)+\alpha_1^2][s(s+\delta_2)+\alpha_2^2]
+g^2(s+\delta_1)(s+\delta_2),
\end{equation}We let $\Delta=\delta_1-\delta_2>0$. At an EP4, all four eigenvalues coincide at $-\beta$, where $\beta>0$. Matching $P(s)=(s+\beta)^4$ gives $\beta=(\delta_1+\delta_2)/4$ and
\begin{align}\label{ir:match-four}
\alpha_1^2+\alpha_2^2+g^2+\delta_1\delta_2=6\beta^2,\,\,\
\alpha_1^2\delta_2+\alpha_2^2\delta_1+4\beta g^2=4\beta^3,\,\,\
\alpha_1^2\alpha_2^2+g^2\delta_1\delta_2=\beta^4.
\end{align}
The first two equations yield
\begin{equation}\label{ir:alphas-four}
\alpha_{1,2}^2=\beta^2+\frac{\Delta^2}{8}-\frac{g^2}{2}
\pm\left(\frac{\beta\Delta}{2}+\frac{2\beta g^2}{\Delta}\right),
\end{equation}
with the upper sign for dot 1. Substitution into the last equation gives
\begin{equation}\label{ir:gcondition}
F(g^2)\equiv
\left(\frac{\Delta^2}{4}-4\beta^2\right)g^4
+\left(\beta^2\Delta^2-\frac{3\Delta^4}{8}\right)g^2
+\frac{\Delta^6}{64}=0.
\end{equation}
Since $0<\Delta<4\beta$, this quadratic equation in $g^2$ has exactly one positive root. A physical EP4 also requires $\alpha_1^2,\alpha_2^2>0$. With all couplings nonzero, the reordered matrix with the convention $(\hat a_1,\hat d_1,\hat d_2,\hat a_2)$ is tridiagonal, and its eigenvector equations determine every component from the first. Its eigenspace is therefore one dimensional, so the quadruple root forms a single Jordan block.
At $\delta_1=3\delta_2$, equivalently $\Delta=2\beta$, these equations give $g^2=\beta^2/3$ and $\alpha_2=0$. For equal widths, the first two matching equations instead force $g=0$. We discard both these cases as they disconnect the chain. At the EP4, the matrix exponential terminates at cubic order:
\begin{equation}\label{ir:four-expand}
e^{\tilde A t}=e^{-\beta t}\sum_{k=0}^3\frac{(\tilde A+\beta\mathds 1)^kt^k}{k!},
\qquad
u_j(t)=e^{-\beta t}(1+\beta t+b_jt^2+c_jt^3).
\end{equation}
Using $(\tilde A^2)_{jj}=-(g^2+\alpha_j^2)$ and $(\tilde A^3)_{jj}=\delta_j\alpha_j^2$ gives
\begin{equation}\label{ir:bcphysical}
b_j=\frac{\beta^2-g^2-\alpha_j^2}{2},
\qquad
c_j=\frac{\beta^3-3\beta(g^2+\alpha_j^2)+\delta_j\alpha_j^2}{6}.
\end{equation}
Substituting Eq.~\eqref{ir:alphas-four},
\begin{align}\label{ir:b-sign}
b_1=-\frac{\delta_1}{2\Delta}\left(g^2+\frac{\Delta^2}{4}\right),\,
b_2=\frac{\delta_2}{2\Delta}\left(g^2+\frac{\Delta^2}{4}\right),\,
c_1=\frac{(\Delta+2\beta)[\Delta^3-4g^2(\Delta+4\beta)]}{96\Delta},\,
c_2=-\frac{(\Delta-2\beta)[\Delta^3+4g^2(4\beta-\Delta)]}{96\Delta}.
%\label{ir:c-sign}
\end{align}which shows that $b_1<0$ and $b_2>0$. Eq.~\eqref{ir:gcondition} further gives
\begin{equation}\label{ir:cproduct}
c_1c_2=-\frac{g^2(\Delta^2-4\beta^2)^2}{576}<0.
\end{equation}
Thus $c_2$ has the sign of $2\beta-\Delta$, and $c_1$ has the opposite sign.
\subsubsection{Counting revivals}
Differentiating the amplitude gives
\begin{equation}\label{ir:four-derivative}
\dot u_j(t)=te^{-\beta t}
\left[-(g^2+\alpha_j^2)+(3c_j-\beta b_j)t-\beta c_jt^2\right].
\end{equation}
Whenever $c_j<0$, the amplitude polynomial has exactly one coefficient sign change, regardless of $b_j$, so Descartes' rule gives exactly one simple positive zero. The derivative bracket has negative constant and positive leading coefficient, hence exactly one positive root. The amplitude therefore crosses zero, reaches a negative minimum, and returns to zero; its square has exactly one revival.
For $1<\delta_1/\delta_2<3$, we have $c_1<0<c_2$. Dot 1 therefore has one revival. For dot 2, $b_2,c_2>0$ and
\begin{equation}\label{ir:monotonic-four}
3c_2-\beta b_2
=-\frac18\left\{
g^2(4\beta-\Delta)
+\Delta\left[\frac{(\Delta-2\beta)^2}{4}+\beta^2\right]
\right\}<0.
\end{equation}
Its amplitude polynomial has positive coefficients, while every coefficient in its derivative bracket is negative. Thus $u_2>0$ and $\dot u_2<0$ for $t>0$, giving monotonic distinguishability.
For $\delta_1/\delta_2>3$, we have $c_2<0<c_1$, so dot 2 has one revival. To establish two revivals for dot 1, first note that
\[
F(\Delta^2/20)=\frac{\Delta^4(16\beta^2-\Delta^2)}{400}>0.
\]
Since $F$ has only one positive root, the EP4 solution satisfies $g^2>\Delta^2/20$. Consequently,
\begin{align}
e^{6\beta/\Delta}u_1(6/\Delta)
=1+\frac{3\beta}{2\Delta}
-\frac{g^2(18\Delta^2+90\beta\Delta+72\beta^2)}{\Delta^4}
<\frac1{10}-\frac{3\beta}{\Delta}
-\frac{18\beta^2}{5\Delta^2}<0,
\end{align}
where the last inequality follows from $\Delta<4\beta$.
The amplitude polynomial starts positive and tends to $+\infty$, so it has a positive zero on each side of $6/\Delta$. Its coefficient signs $+,+,-,+$ permit at most two positive roots by Descartes' rule; hence both zeros are simple and there are no others. The amplitude has a negative minimum between them and a positive maximum after the second zero. These exhaust the two possible positive extrema in Eq.~\eqref{ir:four-derivative}, giving exactly two revivals of $u_1^2$.

\subsection{General result for higher-order non-Markovian EPs}
\label{app:73}
\subsubsection{Polynomial form and initial decay}
Consider $N_s$ dots interacting with $N_b$ independent Lorentzian baths. As before, in the RC description, the number of modes is $m=N_s+N_b$. The geometry of the setup can in principle be arbitrary, except that the dots must not share a common bath. We prove that when the evolution matrix has an EP of order $m$ (i.e., an EP of maximal order), the distinguishability of any observed dot has at most $m-2$ positive-time zeros and at most $m-2$ revival intervals. As before, we remove the common dot energy $\epsilon_d$ and denote the shifted EP eigenvalue by $-\beta$. The diagonal entries of $\tilde A$ are zero on the dots and $-\delta_j$ on the RCs. Since all $m$ eigenvalues equal $-\beta$, taking the trace gives $\beta=\sum_j\delta_j/m>0$, where the sum runs over the RCs. Since the entire matrix forms one EP, $(\tilde{A}+\beta\mathds{1})^m=0$. For a fixed observed dot $j$, the amplitude $u_j(t)$ has the form
\begin{equation}\label{ir:general-amplitude}
u_j(t)=[e^{\tilde{A} t}]_{jj}=e^{-\beta t}p(t),
\qquad
p(t)=\sum_{k=0}^{d}b_kt^k,
\qquad
b_k=\frac{[(\tilde{A}+\beta\mathds{1})^k]_{jj}}{k!}.
\end{equation}
Note that an EP of order $m$ can give rise to a polynomial of degree $m-1$. However, the coefficients can be zero. Here, $d\leq m-1$ is the actual polynomial degree and $b_d\ne0$. Because the observed dot is resonant and undamped, $\tilde{A}_{jj}=0$, so
\[
u_j(0)=1,\qquad \dot u_j(0)=0,
\qquad b_0=1,\qquad b_1=\beta.
\]
Therefore, $d\geq1$.
We also need the initial direction of change of the distinguishability $\mathcal D(t)=|u_j(t)|^2$. Since $\dot u_j(0)=0$ and the coherent couplings satisfy $\tilde{A}_{kj}=-\tilde{A}_{jk}^*$,
\begin{equation}\label{ir:curvature}
\ddot{\mathcal D}(0)
=2\operatorname{Re}(\tilde{A}^2)_{jj}
=-2\sum_{k\ne j}|\tilde{A}_{jk}|^2<0.
\end{equation}
The inequality is strict because a decoupled undamped dot would have eigenvalue zero, incompatible with the EP eigenvalue $-\beta<0$. Thus $\mathcal D(t)$ initially decreases.

\subsubsection{Bound on positive-time zeros}
 The zeros of $\mathcal D(t)$ are precisely the real zeros of $p(t)$. Suppose that this degree-$d$ polynomial had $d$ positive roots $t_1,\ldots,t_d$, counted with multiplicity. These roots would exhaust its degree, giving
\[
p(t)=b_d\prod_{\ell=1}^{d}(t-t_\ell).
\]
Differentiating this product and evaluating at zero yields
\begin{equation}
\frac{p'(0)}{p(0)}
=-\sum_{\ell=1}^{d}\frac1{t_\ell}<0.
\end{equation}
However, the initial conditions give $p'(0)/p(0)=b_1/b_0=\beta>0$. Hence $p(t)$ has at most $d-1$ positive roots counted with multiplicity, and $\mathcal D(t)$ has at most $d-1\leq m-2$ distinct positive-time zeros.

\subsubsection{Bound on revival intervals}
We now count revivals directly, without assuming that each revival follows a zero. Define $\sigma(t)=\dot{\mathcal D}(t)$ for the same fixed pair of initial states. Eq.~\eqref{ir:general-amplitude} gives
\begin{equation}\label{ir:sigma-poly}
\sigma(t)=e^{-2\beta t}R(t),
\qquad
R(t)=\frac{d}{dt}|p(t)|^2-2\beta|p(t)|^2.
\end{equation}
Since $|p(t)|^2$ is a real polynomial of degree $2d$, $R(t)$ also has degree $2d$, with leading coefficient $-2\beta|b_d|^2<0$. Moreover,
\[
R(0)=0,\qquad R'(0)=\ddot{\mathcal D}(0)<0.
\]
Thus $R(t)$ has a simple root at zero and at most $2d-1$ positive roots counted with multiplicity. Its sign, and therefore that of $\sigma(t)$, is negative both immediately after zero and at sufficiently late times.
Each revival, understood as a growth interval separated from other revivals by decay, requires two positive-time sign changes: one from $\sigma<0$ to $\sigma>0$, and another back to $\sigma<0$. If there are $k$ revivals, $R(t)$ must therefore have at least $2k$ positive roots. Consequently,
\begin{align}
2k\leq2d-1,
\qquad
k\leq d-1\leq m-2.
\end{align}
%An isolated stationary point without a sign change does not split a revival into two. 
Together with the zero count above, this proves Eq.~\eqref{ir:bound}.

\section{Escape time and coupled reaction coordinates}

\label{app:esc}

Consider a single dot (site 1) connected to a chain of coupled RCs (sites 2,...,$m$), where site $m$ is damped. The dot and RCs are all resonant such that $A = -i\varepsilon_d\mathds 1  + \Tilde{A}$. We consider the case where the system initially contains a single excitation at site 1 and that all RCs are initially empty and that no initial correlations are present i.e. $\ket{\psi(0)} = \ket{1}$. We also assume that the residual bath is initially empty, so it does not inject excitations into the chain. We then have
\begin{equation}
    \ket{\psi(t)} = D(t)\ket{1} = \sum_{j}^m D_{j1}(t)\ket{j},
\end{equation}
with the probability that the single excitation remains in the system given by
\begin{equation}
    \mathcal{P}(t) = \sum_{j}^m \abs{D_{j1}}^2 = \bra{1}D^\dagger(t)D(t)\ket{1}.
    \label{eq:escp}
\end{equation}
Since we assume that the jump is irrevocable, $\mathcal{P}$ is equal to the probability of the escape time, $T$, being larger than $t$. Hence, the negative time derivative of $\mathcal{P}$ gives the probability density for the escape time
\begin{equation}
    w(t) = -\dot{\mathcal{P}}(t).
\end{equation}
Differentiating Eq.~\eqref{eq:escp} gives
\begin{equation}
    \dot{\mathcal{P}}(t) = \bra{1}D^{\dagger}(t)(A^\dagger+A)D(t)\ket{1} = \bra{1}D^{\dagger}(t)(-2m\beta \ketbra{m}{m})D(t)\ket{1} = -2m\beta \abs{D_{m1}(t)}^2, 
\end{equation}
using $D(t)=e^{At}$ and $A+A^\dagger = -2m\beta \ketbra{m}{m}$, taking the amplitude damping rate out of RC site $m$ to be $m\beta$. Consider now the decomposition $\Tilde{A} = -\beta\mathds 1 + N$. At the maximal EP all eigenvalues of $\tilde A$ are equal to $-\beta$, the Jordan form of $\tilde{A}$ implies $N$ is nilpotent of index $m$, $N^m = 0, N^{m-1} \neq 0$. We can then write
\begin{equation}
    \abs{D_{m1}(t)}^2 = e^{-2\beta t}\abs{(e^{Nt})_{m1}}^2 = e^{-2\beta t}\abs{\sum_k \frac{(t)^k}{k!}(N^k)_{m1}}^2.
\end{equation}
Writing out the product above
\begin{equation}
    (N^k)_{m1} = \sum_{i_1,i_2,...,i_{k-1}}N_{m,i_{k-1}}...N_{i_2,i_1}N_{i_1,1},
\end{equation}
this sum can geometrically be seen as the sum of products of matrix elements along paths connecting node $m$ and $1$ in exactly $k$ steps. It follows that it's zero for $k < m-1$ and if $k = m-1$ the only non-vanishing term is
\begin{equation}
    (N^{m-1})_{m1} = N_{m,m-1}...N_{3,2}N_{2,1} = (-i)^{m-1}\prod_{j=1}^{m-1}g_j.
\end{equation}
Such that
\begin{equation}
    w(t) = \frac{2m\beta \prod_{j=1}^{m-1}g_j^2}{((m-1)!)^2}e^{-2\beta t}t^{2m-2} = C e^{-2\beta t}t^{2m-2}.
\end{equation}
Since $w(t)$ is normalized, $\int_{0}^\infty w(t) dt = 1$, 
\begin{equation}
   C = \frac{2m\beta \prod_{j=1}^{m-1}g_j^2}{(m-1)!(m-1)!} = \frac{(2\beta)^{2m-1}}{(2m-2)!}.
\end{equation} 
The resulting distribution for the escape time $T$ is an Erlang distribution with $k=2m-1, \lambda = 2\beta$ giving
\begin{equation}
    \braket{T} = \frac{2m-1}{2\beta}, \hspace{4pt} \text{Var}(T) = \frac{2m-1}{4\beta^2}.
\end{equation}
Now consider the eigenvalues of $\tilde{A}$, given by $\lambda_k = -r_k + i \omega_k$ and define $\gamma = \min r_k$. Using the following trace relations for $\Tilde{A}$
\begin{equation}
\operatorname{tr}(\tilde A)=-m\beta,\qquad
\operatorname{tr}(\tilde A^2)=(m\beta)^2-2\sum_{j=1}^{m-1}g_j^2.
\end{equation}
Equating the real parts of both relations we have
\begin{equation}
   \sum_k r_k=m\beta,\qquad \sum_k(r_k^2-\omega_k^2)=(m\beta)^2-2\sum_{j=1}^{m-1}g_j^2,
\end{equation}combining the expressions and using that $r_k r_j \geq \gamma^2$ we get
\begin{equation}
  \gamma\leq\sqrt{\frac{2\sum_{j=1}^{m-1}g_j^2}{m(m-1)}}.
\end{equation}
The bound is saturated when 
\begin{equation}
    \sum_{i<j}(r_ir_j-\gamma^2)+\frac12\sum_k\omega_k^2=0,
\end{equation}which is only possible for $\omega_k = 0$ and $r_k = \gamma \hspace{3pt} \forall k$, i.e. when all eigenvalues coincide. Now we show that there is only one independent eigenvector corresponding to this eigenvalue. Consider the vector that satisfies the eigenvalue equation $\tilde A \vec{v} = -\gamma \vec v$, the first component $v_1$ is related to $v_2$ by $-ig_1v_2=-\gamma v_1$, similarly the second component is constrained by $-ig_2v_3-ig_1v_1=-\gamma v_2$. Continuing down the chain we see that all $v_i$ are determined by one parameter $v_1$, there is then only one eigenvector corresponding to the eigenvalue $-\gamma$ with algebraic multiplicity $m$, showing that this is an EP of order $m$. For the connected chain ($g_j\neq0$) it similarly follows that every left eigenvector has a nonzero component at site 1. Since the initial excitation overlaps with the slowest-decaying mode we have 
\begin{equation}
    \lim_{t\to\infty}[-\log\mathcal P(t)]/t=2\gamma,
\end{equation} the polynomial factors from Jordan blocks do not affect this limit. Defining $\mathcal{P}(T_\varepsilon) = \varepsilon$ implies $T_\varepsilon/\log(1/\varepsilon)\to1/(2\gamma)$, and hence
$\lim_{\varepsilon\to0}T_\varepsilon^{\mathrm{Max,EP}}/T_\varepsilon^{\mathrm{non\text{-}EP}}=\gamma_{\mathrm{non\text{-}EP}}/\beta_{\mathrm{EP}}<1$
for any fixed competing configuration that is not a maximal EP and has the same chain length ($m$) and sum of squared coherent couplings ($\sum_{j=1}^{m-1}g_j^2$).

\end{document}